\documentclass[12pt]{article}
\usepackage{latexsym}
\usepackage{epsfig,amssymb,euscript,slashed}
\usepackage{amsmath}
\usepackage{cite} 
\usepackage{array,calc,epsfig}
\usepackage{bbm}
\usepackage{fancybox}
\usepackage[linktocpage]{hyperref}

\def\be{\begin{equation}}
\def\ee{\end{equation}}
\def\bseq{\begin{subequations}}
\def\eseq{\end{subequations}}

\def\bea{\begin{eqnarray}}
\def\eea{\end{eqnarray}}
\newcommand\bbone{\ensuremath{\mathbbm{1}}}
\newcommand{\ul}{\underline}
\def\bseq{\begin{subequations}}
\def\eseq{\end{subequations}}
\def\beq{\begin{equation}}
\def\eeq{\end{equation}}

\numberwithin{equation}{section} 
\usepackage[]{graphicx}

\def\d {{\rm d}}

\def\calb{{\cal B}}
\def\calc{{\cal C}}
\def\cald{{\cal D}}
\def\cale{{\cal E}}

\def\calh{{\cal H}}

\def\call {{\cal L}}

\def\calr         {{\cal R}}

\def\del          {\partial}

\def\ii {{\rm i}}

\def\sqr#1#2{{\vcenter{\vbox{\hrule height.#2pt
 \hbox{\vrule width.#2pt height#1pt \kern#1pt \vrule width.#2pt}\hrule
 height.#2pt}}}}

\def\d{\text{d}}

\def\slashchar#1{\setbox0=\hbox{$#1$}           
\dimen0=\wd0                                 
\setbox1=\hbox{/} \dimen1=\wd1               
\ifdim\dimen0>\dimen1                        
\rlap{\hbox to \dimen0{\hfil/\hfil}}      
#1                                        
\else                                        
\rlap{\hbox to \dimen1{\hfil$#1$\hfil}}   
/                                         
\fi}

\begin{document}
\font\cmss=cmss10 \font\cmsss=cmss10 at 7pt

\begin{flushright}{\scriptsize DFPD-13-TH-08 \\  \scriptsize QMUL-PH-13-07}
\end{flushright}
\hfill
\vspace{18pt}
\begin{center}
{\Large \textbf{6D microstate geometries from 10D structures}}
\end{center}

\vspace{8pt}
\begin{center}
{\textsl{ Stefano Giusto$^{\,a, b}$, Luca Martucci$^{\,a, b}$, Michela Petrini$^{\,c}$ and Rodolfo Russo$^{\,d}$}}

\vspace{1cm}

\textit{\small ${}^a$ Dipartimento di Fisica ed Astronomia ``Galileo Galilei",  Universit\`a di Padova,\\Via Marzolo 8, 35131 Padova, Italy} \\  \vspace{6pt}

\textit{\small ${}^b$ I.N.F.N. Sezione di Padova,
Via Marzolo 8, 35131 Padova, Italy}\\
\vspace{6pt}

\textit{\small ${}^c$ Laboratoire de Physique Th\'eorique et Hautes Energies,\\ Universit\'e Pierre et Marie Curie, CNRS UMR 7589, F-75252 Paris Cedex 05, France}\\
\vspace{6pt}

\textit{\small ${}^d$ Centre for Research in String Theory, School of Physics and Astronomy\\
Queen Mary University of London,
Mile End Road, London, E1 4NS,
United Kingdom}

\end{center}

\vspace{12pt}

\begin{center}
\textbf{Abstract}
\end{center}

\vspace{4pt} {\small
\noindent 
We use the formalism of Generalised Geometry to characterise
in general the supersymmetric backgrounds in type II supergravity that
have a null Killing vector. We then specify this analysis to configurations that 
preserve the same supersymmetries as the D1-D5-P system compactified 
on a four-manifold. We give a set of equations on the forms
defining the supergravity background that are equivalent to the
supersymmetry constraints and the equations of motion. This study is
motivated by the search of new microstate geometries for the
D1-D5-P black hole. As an example, we rewrite the linearised
three-charge solution of arXiv:hep-th/0311092 in our formalism and
show how to extend it to a non-linear, regular and asymptotically flat
configuration. }

\vspace{1cm}


\thispagestyle{empty}

\vfill
\vskip 5.mm
\hrule width 5.cm
\vskip 2.mm
{
\noindent  {\scriptsize e-mails:  {\tt stefano.giusto@pd.infn.it, luca.martucci@pd.infn.it, petrini@lpthe.jussieu.fr, r.russo@qmul.ac.uk} }
}

\newpage

\setcounter{footnote}{0}

\tableofcontents

\newpage



\section{Introduction}

Supersymmetric solutions of type II supergravities play a central
role in many different areas of string theory, including compactifications with interesting phenomenological
properties, the study of holographic gauge/string dualities and the construction 
of geometries with the same charges as black holes.  
Supersymmetry is also a powerful tool in finding explicit solutions, since it allows to solve first order equations
instead of the full set of equations of motion.

In all the examples mentioned above, the supergravity solutions are characterised by the presence of non-trivial Ramond-Ramond
(RR) fields, which  makes it more complicated to solve the Killing spinor equations. In the past years, 
several techniques, from G-structures to Generalised Geometry,  have been developed to simplify
the analysis of the Killing spinor equations in presence of fluxes and make manifest the constraints  that supersymmetry puts
on the background geometry.
In particular, for compactifications to four dimensions, the formalism of Generalised Geometry \cite{Hitchin:2004ut,Gualtieri:2003dx} provides a very elegant tool
for  classifying  the internal geometries  as well as constructing explicit solutions \cite{Grana:2004bg,Grana:2005sn} and studying the physics of D-branes~\cite{Martucci:2005ht,Martucci:2006ij}. The idea is that it is possible to give a completely equivalent
formulation of the Killing spinor equations as a set of differential equations on two polyforms defined on the internal six-dimensional manifold.
The two polyforms are constructed by tensoring the supersymmetry parameters on the six-dimensional space. See \cite{Gauntlett:2002fz,Gauntlett:2002sc,Gauntlett:2003wb,HackettJones:2004yi,Saffin:2004ar} for previous works reformulating Killing spinor equations in terms of differential forms.

Recently this approach has been generalised in \cite{Tomasiello:2011eb} to the analysis of generic ten-dimensional backgrounds in type II
supergravity.  Again, the supersymmetry constraints are rephrased in terms of forms, built out of the ten-dimensional supersymmetry parameters.
In type II supergravities in ten dimensions,  supersymmetric backgrounds are characterised by the presence of a Killing vector, which can be either 
time-like or null.  In this paper we will make use of the formalism developed in~\cite{Tomasiello:2011eb} to study in detail backgrounds with
a null Killing vector.  

This analysis is  motivated by the construction of new microstate
geometries for the 5D Strominger-Vafa (SV) black hole~\cite{Strominger:1996sh,Callan:1996dv}.  The SV black hole has a  realisation in type IIB string theory compactified on a circle of radius $R\gg
\sqrt{\alpha'}$ times a string-sized four dimensional space. For concreteness, in this paper we focus on the $T^4$ case, but our results are valid for internal K3 as well. The SV black hole carries
three charges which, in string theory language, correspond to
D1-branes wrapped on the $S^1$, D5's wrapped on $S^1 \times T^4$ and
left-moving momentum (in our conventions) along the $S^1$. Our main
goal is to set up the framework for building horizon-less solutions that carry the
same charges as the SV black hole and thus are relevant for the so-called ``fuzzball''
conjecture  (see the review
articles~\cite{Mathur:2005zp,Bena:2007kg,Skenderis:2008qn,Mathur:2008nj,Balasubramanian:2008da,Chowdhury:2010ct,Mathur:2012zp}
for a general account of this subject). The first examples of exact and
regular supegravity configurations with three charges were found almost ten years
ago~\cite{Lunin:2004uu,Giusto:2004id,Giusto:2004ip} and
are part of a class of smooth $1/8$-BPS solutions constructed
in~\cite{Bena:2005va,Berglund:2005vb}. 

These geometries are characterised by the presence of a dipole charge for each of 
the three charges; moreover they are smeared over the compact space $S^1 \times
T^4$ (and thus can be studied within a 5D supergravity, see~\cite{Gibbons:2013tqa} for a recent analysis from this point of view) and are further restricted by the assumption that they admit a tri-holomorphic U(1) isometry.

There is evidence that the generic microstate of the SV
black hole does not fall in the class of geometries discussed
above.  First of all, the assumption of a tri-holomorphic U(1) isometry is clearly too
restrictive as it would suppress most of the states already in the two-charge case; this was
confirmed by the countings in~\cite{Bena:2006is,deBoer:2008zn,deBoer:2009un}. 
The assumption of a tri-holomorphic U(1) isometry was relaxed in~\cite{Bena:2010gg}, but a crude counting argument suggested that even those more general solutions could not account for the full three-charge entropy. 

Further evidence that the generic microstate geometry might be more general than \cite{Bena:2005va,Berglund:2005vb}
comes from a complementary analysis of  D1-D5-P systems based on world-sheet techniques \cite{Giusto:2011fy,Giusto:2012jx}.
The basic idea is that string correlators in flat space capture the large distance
behaviour of the gravitational backreaction of each D-brane bound
state.  This approach has been successfully tested for bound states of D1-D5 branes  \cite{Giusto:2009qq,Black:2010uq} where it was shown that 
 disk correlators with one closed string and all open strings  describing the given microstate  reproduce the
dipole charges for the known  two-charge configurations.  The same analysis applied to the three-charge geometries  \cite{Giusto:2012jx}
suggests that the generic SV
microstate is not smeared along the $S^1$ and thus should be described
within a 6D supergravity\footnote{A first example of solution in this
  class was found in~\cite{Ford:2006yb}. The fact that generic microstates should be described by 
 un-smeared six-dimensional geometries was also conjectured in~\cite{Bena:2011uw} by analysing the supersymmetries preserved by D1-D5-P bound states.}. Solutions of minimal 6D supergravity coupled to (at most) one tensor multiplet and carrying the same charges and dipole charges as the~\cite{Bena:2005va,Berglund:2005vb} solutions  
 were studied in~\cite{Gutowski:2003rg,Cariglia:2004kk,Bena:2011dd}.  However, the world-sheet analysis also shows that all fields of
type IIB supergravity are non-trivial for general microstates, and not only those necessary for
describing the dipoles
present in~\cite{Bena:2005va,Berglund:2005vb}. This conclusion is
supported also by arguments based on the dual  D1-D5
CFT~\cite{Kanitscheider:2007wq}. 
Thus we should expect that for the generic D1-D5-P microstates all
type IIB fields are excited and, hence,  cannot be described by 
a restricted ansatz  based on minimal six-dimensional supergravity like the
one discussed in~\cite{Gutowski:2003rg,Cariglia:2004kk,Bena:2011dd}.

The first aim of this paper is to provide the most general ansatz and set of  equations 
that  should describe the microstates of the D1-D5-P system directly in ten dimensions, with only two simplifying assumptions:  we assume that the $T^4$ is rigid,
i.e. that it appears in the 10D geometry as a fixed torus apart from
an overall scaling function, and that the solution is `isotropic' on $T^4$. 
Of course, this does not happen for the
generic microstate, but, as in the two-charge case, there is the
possibility that the class of geometries we consider gives an entropy
that has the correct scaling in the large charge limit.
 Our analysis  yields a set of equations generalizing the results
of~\cite{Gutowski:2003rg,Cariglia:2004kk,Bena:2011dd}.
The presence of the momentum charge implies the existence of a null Killing
vector, while the requirement of supersymmetric D1 and D5 charges is  necessary and
sufficient to ensure that $1/8$ of the total type IIB supersymmetries
are preserved.

The second aim of the paper is to provide an explicit
solution of our system of equations describing  a smooth D1-D5-P microstate geometry that depends non-trivially on the $S^1$ coordinate\footnote{In this sense, the geometry we obtain is a regular solution of a non-minimal 6D supergravity; however, the $S^1$ is part of the 6D space and the D-branes are wrapped on it: thus from our example it is not possible to derive a regular `soliton' of 6D supergravity in the asymptotically Minkowski space $\mathbb{R}^{1,5}$.}. 
A class of such solutions would be provided by the non-linear completion of the perturbative
geometries derived from string theory in~\cite{Giusto:2012jx}. However, for 
sake of simplicity we focus on a different case: we start from the
perturbative 6D solution derived in~\cite{Mathur:2003hj} and show that
it can be uplifted to 10D as a two-charge solution of the general type
constructed in~\cite{Kanitscheider:2007wq}. Then we can follow the
construction of~\cite{Mathur:2003hj}, adding a non-trivial momentum
and extending the solution to an asymptotically flat geometry. The
knowledge of the full 10D equations allows us to perform these steps at
the non-linear level and to find a 10D completion of the solution
discussed in~\cite{Mathur:2003hj}.

\vspace{0.2cm}

The paper is organised as follows. 
In Section \ref{sec:susyeq} we give the complete set of equations one has to solve in order to have a ten-dimensional supersymmetric solution
with a null isometry in type II supergravity. As already mentioned our starting point is the formalism developed in  \cite{Tomasiello:2011eb}.  
We construct  a one-form, one even/odd polyform in IIA/IIB and a pair of four-forms, which are obtained as (sum of) bilinears in the supersymmetry parameters, and then
we give the set of equations on these objects that are equivalent to the Killing spinor equations.  In doing that, we derive a new set of equations, which appear simpler than those in \cite{Tomasiello:2011eb}, but only holds for backgrounds with a null isometry.
Morevoer, we show that, in order to have a full solution of all supergravity equations of motion, we need also to impose the component of Einstein equations
in the null direction.  
Then in  Section \ref{sec:15P} we will specify our equations to the study of a D1-D5-P (marginal) bound states.  We will consider geometries of the form $\mathbb{R}^{1,1}\times Y\times \hat{Y}$, where $Y$ is fibrated on the null direction in $\mathbb{R}^{1,1}$ and $\hat Y$ is eventually identified with the internal $T^4$ (or K3).  Under the assumption that all fields are homogeneous and isotropic along $\hat{Y}$, we derive  the most
general ansatz for the type IIB fields describing such systems and the equations that such fields must satisfy in order to give a solution.
These first two sections can be skipped by readers who are more interested in the construction of explicit geometries.  In order to make the second part of the paper self-contained and consistent with the notation used in the existing literature on microstate geometries, we
collect in the Appendix~\ref{app:summary} a summary of the equations
relevant for the concrete applications to the SV miscrostates.
In Section~\ref{solutionexample}, we briefly review the perturbative
configuration of~\cite{Mathur:2003hj} and discuss its embedding in the 10D 
ansatz  derived in Section \ref{sec:15P}. At this point it is easy to
write a full regular non-linear solution in the near-horizon
region. The extension to an asymptotically flat configuration is
trickier: here we present the complete solution but postpone the
derivation and a more detailed discussion of its physical properties
to a forthcoming paper.
Conventions and details of the derivations of Sections \ref{sec:susyeq} and \ref{sec:15P} can be found in the Appendices.


\section{Null supersymmetric vacua: general discussion}
\label{sec:susyeq}

The study of type II flux backgrounds has shown that, rather than dealing directly with spinorial equations, 
 it is often convenient to rewrite the supersymmetry conditions as differential equations on a set of
forms on the compactification manifold.  In particular, in \cite {Tomasiello:2011eb} this approach has been applied to generic 
backgrounds in ten dimensions.  As we will review in this section, the idea is that, starting from the pair of supersymmetry parameters  $\epsilon_1$ and  $\epsilon_2$,  
one can build a set of forms  which can be used to rewrite the  ten-dimensional supersymmetry conditions. 

Ten dimensional supersymmetric solutions in type II supergravity  are characterised by the existence of a 
Killing vector $K$, which can be either time-like or null.  In this paper we will assume a null $K$. We will re-examine the supersymmetry conditions given
in \cite{Tomasiello:2011eb} and propose an alternative formulation of some of the supersymmetry constraints.  

For generic ten-dimensional backgrounds the supersymmetry constraints are not equivalent to the full set of equations of motion.  For the case of $K$ null, we 
show that,  in order to get a proper supergravity solution,  the supersymmetry constraints, the Bianchi identities for the NSNS flux and for
all the RR fluxes of the democratic formalism, summarized in (\ref{BI10}), have to be supplemented by the $vv$ component of Einstein equation.
\subsection{Supersymmetry and geometrical structures}

In ten dimensions, a supersymmetric vacuum  of type  II supergravity  is  characterised by a pair of Majorana-Weyl spinors, 
$\epsilon_1$ and  $\epsilon_2$, satisfying the Killing spinor equations.  We take their chirality to be
\beq
\Gamma_{(10)} \epsilon_1 = \epsilon_1  \qquad\,,  \qquad \Gamma_{(10)} \epsilon_2 =  \mp \epsilon_2  \, ,
\eeq
where the minus (plus) sign is for IIA (IIB). We have introduced the ten-dimensional  chiral operator $\Gamma_{(10)}=\Gamma^{\ul{0\ldots 9}}$. Here and in the following the 
uderline  denotes flat indices.  

\vspace{0.2cm}

Form the spinor bilinears, we can construct a vector $K$  \cite{Tomasiello:2011eb}
\be
K =   -\frac12(\bar{\epsilon}_1 \Gamma^M \epsilon_1 +\bar{\epsilon}_2 \Gamma^M \epsilon_2)\del_M \, , 
\ee
which can be either time-like or null.  It can be shown \cite{HackettJones:2004yi, Saffin:2004ar} that  $K$ is a Killing vector:
\be\label{killingmetric}
\call_Kg_{(10)}=0 \, . 
\ee

In addition to $K$, one can construct other useful tensorial objects which characterise the geometry. A 1-form
\be\label{defchi}
\chi=-\frac12(\bar{\epsilon}_1 \Gamma_M \epsilon_1 -\bar{\epsilon}_2 \Gamma_M \epsilon_2)\d X^M
\ee
and a polyform
\be\label{defpsi}
\Psi=-32\,\epsilon_1\otimes \bar\epsilon_2\Gamma_{(10)}=\sum_{k}\frac{1}{k!}\bar\epsilon_1\Gamma_{M_1\ldots M_k}\epsilon_2\, \d X^{M_1}\wedge \ldots \wedge \d X^{M_k}\,,
\ee
where $k=$ is even/odd in IIA/IIB. We use conventions which are slightly different from those of \cite{Tomasiello:2011eb}, since more `natural'  for studying D-branes \cite{Martucci:2011dn}\footnote{The dictionary is the following: $H_{\rm here}=H_{\rm there}$, $F^{\rm IIA}_{\rm here}=-F^{\rm IIA}_{\rm there}$,  $F^{\rm IIB}_{\rm here}=F^{\rm IIB}_{\rm there}$, $K_{\rm here}=-32K_{\rm there}$, $\chi_{\rm here}=-32\tilde K_{\rm there}$, $\Psi^{\rm IIB}_{\rm here}=32\Phi^{\rm IIB}_{\rm there}$,  $\Psi^{\rm IIA}_{\rm here}=-32\Phi^{\rm IIA}_{\rm there}$.}.

\vspace{0.2cm}

One can then show that the ordinary spinorial supersymmetry conditions, which are obtained by setting to zero the supersymmetry variation of the type II gravitini and dilatini, imply the following differential conditions for $\chi$ and
$\Psi$ \cite{Tomasiello:2011eb}
\bseq\label{gensusy}
\begin{align}
& \d\chi=\iota_K  H \, , \label{gensusya}\\
& \d_H(e^{-\phi}\Psi) =\iota_K F+\chi\wedge F \, , \label{gensusyb}
\end{align}
\eseq
where $\d_H$ denotes the twisted exterior derivative 
\be
\d_H\equiv \d -H\wedge \, , 
\ee
$\phi$ is the dilaton, $H$ is the Neveu-Schwarz three-form, and  $F$, following the democratic formalism,   denotes the sum of all Ramond-Ramond field strengths
\beq
\label{RRdef}
F = \sum_k F_{k} \, , 
\eeq
with  $k$ even  (from 0 to 10) in IIA and odd (from 1 to 9) in IIB. The redundant degrees of freedom  in $F$ are eliminated by the self-duality constraint
\beq
\label{10dselfdbis}
F =  \ast \lambda(F)  \, ,
\eeq
where $\lambda$ acts on forms by inverting the order of the indices: on a form $F_k$ of rank $k$ it acts as $\lambda(F_k)=(-)^{\frac{k(k-1)}{2}}F_k$. 
We assume that the fluxes satisfy the proper the Bianchi
identities 
\beq
\label{BI10}
\d H = 0\qquad\,,\qquad\d_H F = 0  \, . 
\eeq
More details on our conventions are given in Appendix \ref{app:conv}.  

\vspace{0.2cm} 

The form equations  \eqref{gensusy} and the condition \eqref{killingmetric} 
can be used to show that $K$ is actually a symmetry of the
full solution \cite{Tomasiello:2011eb}
\be
 \mathcal{L}_K  H= \mathcal{L}_K F =    \mathcal{L}_K \phi  =  \mathcal{L}_K \Psi  = 0 \, . 
\ee

\vspace{0.2cm}

As explained in \cite{Tomasiello:2011eb}, the equations (\ref{killingmetric}) and (\ref{gensusy}) in general contain  less information than the  complete set of 
Killing spinor equations. In other words, they are necessary but not sufficient for having a supersymmetric configuration and must be supplemented by complementary supersymmetry conditions. In  \cite{Tomasiello:2011eb} a possible way of writing such complementary conditions in terms $\Psi$ and other geometrical data was proposed -- see eqs.\,(3.1c-3.1d) therein. However these extra equations are  quite cumbersome to manipulate when looking for explicit solutions. In the following, restricting to the case of $K$ null,  we propose another set of complementary conditions which might be easier to work with.

\subsection{The null $K$ case}

Let us  assume that $K$ is null. We define a light-cone coordinate $u$ such that 
\be
K=\frac{\del}{\del u} \,.
\ee
As explained for instance in \cite{Gutowski:2003rg}, one can introduce a second coordinate $v$ and write the (string frame) metric as
\be
\d s^2_{(10)} = -2 e^{2D}(\d v+\beta)\,\Big[\d u+\omega + W(\d v+\beta)\Big]+\d s^2_X\,,
\label{metric}
\ee 
where $\d s^2_X=g_{ab}(v,x)\d x^a\d x^b$ is a metric on an 8-dimensional space $X$, 
 $\omega=\omega_a(v,x)\d x^a$ and $\beta=\beta_a(v,x)\d x^a$ are one-forms on $X$. In the applications to the D1-D5-P system we will identify
\be\label{timelike}
u= \frac{1}{\sqrt{2}}(t-y)\qquad\,,\qquad v= \frac{1}{\sqrt{2}}(t+y)\,,
\ee
where $t$ and $y$ label time and the $S^1$. All the fields in the metric can  depend in principle on $v$ and the eight-dimensional coordinates
 $x^a$.  We choose the following vielbeine and covielbeine 
\be
\begin{array}{lcl}
E^{\ul u}=\d u+{\omega }+W(\d v+\beta) \, ,  & \qquad  \quad & E_{\ul u}=\frac{\partial}{\partial u}  \, , \\
E^{\ul v}= e^{2D} (\d v+\beta) \, , & \qquad  & E_{\ul v}= e^{-2D}\Big(\frac{\partial}{\partial v}-W\frac{\partial}{\partial u}\Big) \, ,  \\
E^{\ul {a}}= e^{\ul a}  \, , & \qquad  & E_{\ul {a}}= e^b_{\ul a}\Big(\frac{\del}{\del x^b}-\omega_b\,\frac{\partial}{\partial u}-\beta_b\,\frac{\partial}{\partial v}\Big) \, ,
\end{array}
\eeq
such that $\d s^2_X = \delta_{\ul{ab}}e^{\ul a}\,e^{\ul b}$ and $\d s^2_{(10)}=-2 E^{\ul u}E^{\ul v}+\delta_{\ul{ab}}E^{\ul a}E^{\ul b}$ with, as usual, $e^{\ul a} \, e_{\ul b}=\delta^{\ul a}_{\ul b}$ and $E^{\ul M}  E_{\ul N}=\delta^{\ul M}_{\ul N}$. 

\vspace{0.2cm}

The NSNS and RR field strengths can be split in the following way 
\be\label{fluxdec}
\begin{aligned}
H&=h+E^{\ul u}\wedge h_{\ul{u}}+E^{\ul v}\wedge h_{\ul{v}}+E^{\ul u}\wedge E^{\ul v}\wedge h_{\ul{uv}} \, , \\
F&=f+E^{\ul u}\wedge f_{\ul{u}}+E^{\ul v}\wedge f_{\ul{v}}+E^{\ul u}\wedge E^{\ul v}\wedge f_{\ul{uv}} \ , 
\end{aligned}
\ee
where all the forms $h_{\ldots}$ and $f_{\ldots}$ have components along the 8-dimensional coframe $\d x^a$ (or $e^{\ul a}$) only. 
The ten-dimensional self-duality  of the RR fields (\ref{10dselfdbis})  translates into the following 8-dimensional  relations
\beq
\label{selfdual}
\begin{aligned}
& *_8\lambda(f)=f_{\ul{uv}}  \, , \qquad  &  *_8\lambda(f_{\ul{uv}}) =f   \, , \\
& *_8\lambda(f_{\ul u})=-f_{\ul u}\, ,  \qquad &  *_8\lambda(f_{\ul v})=f_{\ul v} \, . 
\end{aligned}
\eeq

\vspace{0.2cm}

Notice that this parametrization, and in particular the choice of $u$ and $v$, is not unique. For fixed $x^a$ we have the following mixed diffeomorphisms  
\be\label{mixediff}
u\rightarrow u+U(v,x)\quad\,,\quad v\rightarrow v+V(x)\quad\,,\quad  x^a\rightarrow x^a \, , 
\ee
which preserve the ansatz (\ref{metric}) if
\be
\omega\rightarrow \omega-\d_XU+\dot U\,\beta\quad\,,\quad\beta\rightarrow \beta-\d_XV\quad\,,\quad W\rightarrow W-\dot U \, , 
\ee
with  $\dot U\equiv \frac{\d U}{\d v}$.  Since we know that $K$ is  a symmetry of the background,  
the forms on $X$  do not depend on $u$ but can in general depend on $v$.   On such forms
 the exterior derivative $\d_X=\d x^a\wedge \partial_a$ is not covariant under (\ref{mixediff}), but can be naturally replaced by 
the modified exterior derivative:
\be
\cald\equiv \d_X-\beta\wedge \frac{\d}{\d v} \, . 
\ee

\vspace{0.2cm}

$K$ being null also restricts the form of the supersymmetric structures . First of all,  we can choose  the ten-dimensional gamma matrices to be 
\be\label{gamma2+8}
\Gamma^{\ul{u}}=-\sqrt{2}\left(\begin{array}{cc} 0 & 0 \\ 1 & 0 
\end{array}\right)\otimes \gamma_{(8)} \quad\,,\quad \Gamma^{\ul{v}}=\sqrt{2}\left(\begin{array}{cc} 0 & 1 \\ 0 & 0 
\end{array}\right)\otimes \gamma_{(8)}\quad\,,\quad \Gamma^{\ul a}=\bbone\otimes \gamma^{\ul a} \ , 
\ee
where $\gamma^{\ul{a}}$ are eight-dimensional gamma matrices associated with the manifold $X$ and $\gamma_{(8)}\equiv \gamma^{\ul{1\ldots 8}}$ is the eight-dimensional chiral operator. 

\vspace{0.2cm}

The fact that $K\equiv E_{\ul u}$ is null implies that the two supersymmetry parameters $\epsilon_1$ and $\epsilon_2$ can be written as\footnote{In general one can split the type II  supersymmetry parameters as
\be
\epsilon_I = \left(\begin{array}{c} 1 \\ 0 \end{array}\right)\otimes \eta_I + \left(\begin{array}{c} 0 \\ 1 \end{array}\right)\otimes \tilde\eta_I  \nonumber \, .
\ee
with $I=1,2$. If $K$ is null,   $\bar\epsilon_I \Gamma^M\epsilon_I \del_M=\bar\epsilon_I \Gamma^{\ul{u}}\epsilon_I  E_{\ul{u}}$ for both $I=1,2$ and then
\be
\epsilon^\dagger_I \Gamma^{\ul{0}}\Gamma^{\ul v}\epsilon_I \sim \epsilon^\dagger_I (-\bbone+\Gamma^{\ul{01}})\epsilon_I = \epsilon^\dagger_I  [(-\bbone_2+\sigma_3)\otimes\bbone]\epsilon_I =0 \, ,  \nonumber
\ee
which requires $\tilde\eta_I=0$.} 
\be\label{ansatz1}
\epsilon_I =\left(\begin{array}{c} 1 \\ 0 \end{array}\right)\otimes \eta_I  \qquad \qquad (I = 1,2) \, , 
\ee
where $\eta_i$  are real eight-dimensional spinors of positive chirality.

By further imposing the normalization condition $K=\del_u$ we get  
\be\label{normspinor}
\frac{1}{\sqrt{2}}(\eta_1^\dagger\eta_1+\eta_2^\dagger\eta_2)=1 \,  , 
\ee
so that we can parametrize
\be\label{normspin}
\|\eta_1\|^2\equiv \eta_1^\dagger\eta_1=\sqrt{2}\, \sin^2\theta\quad~~\,,\quad~~ \|\eta_2\|^2\equiv \eta_2^\dagger\eta_2=\sqrt{2}\, \cos^2\theta \, . 
\ee

\vspace{0.2cm}

By using this restricted form of the Killing spinors and the gamma matrices given in  (\ref{gamma2+8}), it is not difficult to see that the one-form $\chi$ defined in (\ref{defchi}) reduces to
\be\label{1form}
\chi=\cos2\theta\, e^{2D} (\d v+\beta) \, . 
\ee
On the other hand, the polyform $\Psi$  defined in \eqref{defpsi} becomes
\be\label{splitpsi}
\Psi=   \sqrt{2}\,e^{2D} \sin 2 \theta  ( \d v+\beta) \wedge \Phi \, , 
\ee
where $\Phi$ is a polyform on $X$
\be\label{Phi}
\Phi=\sum_{k}\,\frac{1}{k!}\,\eta_1^\dagger\gamma_{a_1\ldots a_k}\eta_2\,\d x^{a_1}\wedge\ldots\wedge\d x^{a_k} 
\ee
which is odd/even in IIA/IIB.

\subsubsection{The missing supersymmetry conditions}
\label{missingsusy}
As shown in  \cite{Tomasiello:2011eb}, equations (\ref{killingmetric}) and (\ref{gensusy})  contain less information than the original Killing spinor equations.
As we discuss more in detail in Appendix \ref{app:extrasusy}, the analysis of the intrinsic torsion of the Killing spinor equations and of the system \eqref{killingmetric}-\eqref{gensusy} reveals that the missing constraints are exactly provided by the vanishing of the $v$-component
of the gravitino variations\footnote{See Appendix \ref{app:conv} for our convention on the Killing spinor equations.}
\be\label{eqv}
\begin{aligned}
&(\nabla_{\ul v}-\frac14 \iota_{\ul v}H)\epsilon_1+\frac{1}{16}\, e^\phi\, F\, \Gamma_{\ul v}\Gamma_{(10)}\epsilon_2=0 \, , \\
&(\nabla_{\ul v}+\frac14 \iota_{\ul v}H)\epsilon_2-\frac{1}{16}\, e^\phi\, \lambda(F)\Gamma_{\ul v}\epsilon_1=0 \, ,
\end{aligned}
\ee
where $F$ is the sum of all RR-field strengths defined in \eqref{RRdef} and all forms are implicitly contracted by gamma matrices, e.g.\ $F\equiv \sum_k\frac{1}{k!}\, F_{M_1\ldots M_k}\Gamma^{M_1\ldots M_k}$.

 In this section, we discuss how one can rewrite the spinorial equations (\ref{eqv}) as an equivalent set of equations involving just (possibly $v$-dependent) differential forms defined on the internal eight-dimensional space $X$. 

\vspace{0.2cm} 

Let us introduce a pair of four-forms $\Omega^{(1)}$ and $\Omega^{(2)}$ on $X$ as follows
\be
\label{Omega4f}
\Omega^{(1)}_{abcd}=\eta_1^T\gamma_{abcd}\,\eta_1\quad~~~\,,\quad~~~ \Omega^{(2)}_{abcd}=\eta_2^T\gamma_{abcd}\,\eta_2 \, .
\ee
$\Omega^{(1)}$ and $\Omega^{(2)}$ define the two Spin(7) structures associated with the  presence of the two Majorana-Weyl  spinors $\eta_1$ and $\eta_2$. Such Spin(7) structures provide a useful tool to analize the supergravity equations as one can decompose tensors in irreducible representations of these Spin(7) structures. In particular, two-forms  on $X$ contain a component transforming  as the  representations ${\bf 7}$ corresponding to $\Omega^{(1)}$ and $\Omega^{(2)}$, which are  selected  by the following projectors
\be
\begin{aligned}
(P^{(1)}_{\bf 7})_{ab}{}^{cd}&=\frac14\Big(\delta^{[c}_{[a}\delta^{d]}_{b]}-\frac1{2\sqrt{2}\sin^2\theta}\Omega^{(1)}_{ab}{}^{cd}\Big) \, , \\
(P^{(2)}_{\bf 7})_{ab}{}^{cd}&=\frac14\Big(\delta^{[c}_{[a}\delta^{d]}_{b]}-\frac1{2\sqrt{2}\cos^2\theta}\Omega^{(2)}_{ab}{}^{cd}\Big) \,  . \\
\end{aligned}
\ee

Then, as shown in details in Appendix \ref{app:extrasusy}, the two spinorial equations \eqref{eqv} can be recast into the following equivalent set of equations
\begin{subequations}\label{alteqs}
\begin{align}
\label{alteqsa}
\frac{\d}{\d v}(\cos 2\theta) =&  \frac{\sqrt{2}}4e^{2D+\phi} f_{\ul v}\cdot\Phi   \, ,  \\
\label{alteqsb}
\frac1{\sqrt{2}}e^\phi\, (f\cdot\iota_a\Phi)\,\d x^a =&  e^{-2D}\cald e^{2D}-\dot\beta+\cos2\theta\, h_{\ul{uv}}  \, ,  \\
\label{alteqsc}
\frac1{\sqrt{2}}e^\phi\, (\iota_a f\cdot\Phi)\,\d x^a =& -h_{\ul{uv}}-\cos2\theta\,(e^{-2D}\cald e^{2D}-\dot\beta) \, ,  \\
\label{alteqd}
 \iota_{[a}\Omega^{(1)}\,\cdot \,\frac{\d}{\d v}\big(\iota_{b]}\Omega^{(1)}\big) =& -16\sin^4\theta\,\, e^{2D}(P^{(1)}_{\bf 7})_{ab}{}^{cd}(\cald\omega+W\cald\beta-h_{\ul v})_{cd}\cr
&  +\sqrt{2}\sin^2\theta\, e^{2D+\phi}\Phi\cdot ( \gamma_{ab} f_{\ul v}) \, ,  \\
\label{alteqe}
\iota_{[a}\Omega^{(2)}\,\cdot \,\frac{\d}{\d v}\big(\iota_{b]}\Omega^{(2)}\big) =&-16\cos^4\theta\,\, e^{2D}(P^{(2)}_{\bf 7})_{ab}{}^{cd}(\cald\omega+W\cald\beta+h_{\ul v})_{cd}\cr
&  +\sqrt{2}\cos^2\theta\, e^{2D+\phi}\Phi\cdot ( f_{\ul v}\gamma_{ab}) \, .
\end{align}
\end{subequations}
In the above equations, the gamma matices $\gamma_a$ must be considered as the corresponding operators under Clifford map: if $\omega_p$ is a $p$-form, then
\be
\begin{aligned}
 &\gamma_a\omega_p\equiv\iota_a \omega_p +g_{ab}\d x^b\wedge \omega_p \, , \cr
 &\omega_p \gamma_a\equiv(-)^{p-1}(\iota_a\omega_p-g_{ab}\d x^b\wedge \omega_p) \, .
\end{aligned}
\ee
Furthermore, as usual, $\gamma_{ab}\equiv\gamma_{[a}\gamma_{b]}$. Notice that in (\ref{alteqd}) and (\ref{alteqd})
only the ${\bf 7}$ components (with respect  to $\Omega^{(1)}$ and $\Omega^{(2)}$ respectively) of the antysimmetric two-tensor are non-trivial. 

Equations  \eqref{alteqsa}-\eqref{alteqe} provide, for the case of $K$ null,  a set of constraints, alternative to Eqs.~(3.1c)-(3.1d) of \cite{Tomasiello:2011eb}, 
 that need to be imposed  in addition to (\ref{killingmetric}) and (\ref{gensusy})   in order to have a supersymmetric configuration.

\subsubsection{Einstein and $B$-field equations}
\label{intnull}

It is well known that supersymmetry does not generically imply the complete set of equations of motion. 
The question is then which is the  minimal set of equations one has to solve in order to be sure to have a solution of the full system  of equations of motion.  The general strategy is
to first impose the supersymmetry and Bianchi identities for the fluxes, and then to check whether these imply the flux  and dilaton equations of motion and Einstein equations. 

In ten dimensions,   one has again to distinguish between backgrounds with null or time-like Killing vectors.  We show in Appendix   \ref{app:intnull} that for $K$ null, one has to 
solve the following set of equations:  the supersymmetry constraints, the Bianchi identities (\ref{BI10}) and the  $\ul{vv}$
component of the Einstein equations
\be
 \calr_{\ul{vv}}+2\nabla_{\ul v}\nabla_{\ul v}\phi-\frac12\iota_{\ul v}H\cdot\iota_{\ul v}H-\frac14 e^{2\phi}\iota_{\ul v}F\cdot\iota_{\ul v}F=0 \, , 
\ee
where $\iota_{\ul v}F\cdot\iota_{\ul v}F=\sum_k\iota_{\ul v}F^{(2k+1)}\cdot\iota_{\ul v}F^{(2k+1)}$ with  $k= 1/2, \ldots, 3/2$ for IIA and $k=0, \ldots 4$ for IIB. 

All other equations, namely the equation of motion for $B$ and the dilaton, and the other component of Einstein equations, are automatically satisfied. \\

With our ansatz for the metric and the fluxes the $\ul{vv}$ component of Einstein equations becomes
\begin{equation}
\begin{aligned}
\label{vvEin}
0 &=  -e^{-4D}\cald_a(e^{2D}L^a)  +  \frac12\, e^{-2D}\beta_aL^a g^{bc}\dot g_{bc}+2e^{-2D}\dot\beta_aL^a \cr
&  + \frac14(\cald\omega+W\cald\beta)_{ab}(\cald\omega+W\cald\beta)^{ab}   -\frac14\, \frac{\d}{\d v}(e^{-4D} g^{ab})\dot g_{ab} \cr
&  -\frac12\, e^{-4D}g^{ab}\ddot{g}_{ab}   +e^{-2D}\frac{\d}{\d v}(e^{-2D}\dot\phi)+e^{-2D}L^a\cald_a\phi \cr
& -\frac{1}{2} h_{\ul v}\cdot h_{\ul v} -\frac{1}{4} e^{2 \phi} f_{\ul{v}}\cdot  f_{\ul{v}}  \, , 
\end{aligned}
\end{equation}
where $h_{\ul v}$ and $f_{\ul v}$ are defined in (\ref{fluxdec}) and the symbol $L$ denotes
\be\label{defLbis}
L = \dot{\omega} + W \dot{\beta} - \cald W \, .
\ee
\subsubsection{Summary of BPS equations for null Killing vectors}
\label{summary}

In the previous sections we derived a set of conditions on the geometric structures $\chi$, $\Psi$, $\Omega^{(1)}$ and $\Omega^{(2)}$ which, for null $K$, are equivalent to the standard Killing spinor
equations of ten-dimensional type II supergravities.  We will summarise them in this section. 

First, in addition to $\call_Kg_{(10)}=0$, we have the equations \eqref{gensusy} \cite{Tomasiello:2011eb}
\begin{subequations}
\begin{align}
\d\chi&=\iota_K H  \, , \label{susysum1a}\\
\d_H(e^{-\phi}\Psi) &=\iota_K F+\chi\wedge F  \, ,  \label{susysum1b}
\end{align}
\end{subequations}
which must be supplemented by the additional constraints\footnote{These equations are a partial rewriting of the constraints discussed in Section \ref{missingsusy}. In particular, (\ref{susysum2a}) and (\ref{susysum2b}) can be obtained from 
\eqref{alteqsa} and \eqref{alteqsb} by using (\ref{susysum1a}).}
\begin{subequations}
\begin{align}
e^{-2D}\cald e^{2D}-\dot\beta+\cald\log(\sin2\theta) =& \frac1{\sqrt{2}\,\sin^2(2\theta)}e^\phi\, (f\cdot\iota_a\Phi)\,\d y^a  \, , \label{susysum2a} \\
\label{susysum2b} 
 \cald_a \cos2\theta =& \frac1{\sqrt{2}}e^\phi\, \iota_a f\cdot\Phi  \,  , \\
\label{susysum2c}
\frac{\d}{\d v}(\cos 2\theta) =&  \frac{\sqrt{2}}4e^{2D+\phi} f_{\ul v}\cdot\Phi   \, ,\\
\label{susysum2d}
 \iota_{[a}\Omega^{(1)}\,\cdot \,\frac{\d}{\d v}\big(\iota_{b]}\Omega^{(1)}\big) =& -16\sin^4\theta\,\, e^{2D}(P^{(1)}_{\bf 7})_{ab}{}^{cd}(\cald\omega+W\cald\beta-h_{\ul v})_{cd}\cr
&  +\sqrt{2}\sin^2\theta\, e^{2D+\phi}\Phi\cdot ( \gamma_{ab} f_{\ul v})  \, , \\
\label{susysum2e}
\iota_{[a}\Omega^{(2)}\,\cdot \,\frac{\d}{\d v}\big(\iota_{b]}\Omega^{(2)}\big) &=-16\cos^4\theta\,\, e^{2D}(P^{(2)}_{\bf 7})_{ab}{}^{cd}(\cald\omega+W\cald\beta+h_{\ul v})_{cd}\cr
 & +\sqrt{2}\cos^2\theta\, e^{2D+\phi}\Phi\cdot ( f_{\ul v}\gamma_{ab})  \, . 
\end{align}
\end{subequations}

Second, in order to have a solution of the full set of equations of motion, we also have to impose the Bianchi identites/eom $\d H=0$ and $\d_HF=0$ and the  $\ul{vv}$ component of Einstein equations
\begin{equation}
\begin{aligned}
\label{sumEins}
0 &=  -e^{-4D}\cald_a(e^{2D}L^a)  +  \frac12\, e^{-2D}\beta_aL^a g^{bc}\dot g_{bc}+2e^{-2D}\dot\beta_aL^a \cr
& + \frac14(\cald\omega+W\cald\beta)_{ab}(\cald\omega+W\cald\beta)^{ab}   -\frac14\, \frac{\d}{\d v}(e^{-4D} g^{ab})\dot g_{ab} \cr
&  -\frac12\, e^{-4D}g^{ab}\ddot{g}_{ab}   +e^{-2D}\frac{\d}{\d v}(e^{-2D}\dot\phi)+e^{-2D}L^a\cald_a\phi \cr
& -\frac{1}{2} h_{\ul v}\cdot h_{\ul v} -\frac{1}{4} e^{2 \phi} f_{\ul{v}}\cdot  f_{\ul{v}}  \, ,
\end{aligned}
\end{equation}
where $h_{\ul v}$, $f_{\ul v}$ and $L$ are defined in (\ref{fluxdec}) and (\ref{defLbis}).

\section{The D1-D5-P geometries}
\label{sec:15P}

In this section we will apply the formalism developed in Section \ref{sec:susyeq}  to study  the full back-reaction of a 
D1-D5-P system, where  the D1-branes wrap an  $S^1 \subset  \mathbb{R}^{1,1}$ while the D5's wrap  $S^1 \times T^4$. 
The momentum represents a left-moving  wave propagating on $S^1$.  
We will therefore consider spaces of the form $\mathbb{R}^{1,1}\times Y\times \hat{Y}$, with metric 
\be
\d s^2_{(10)} = -2 e^{2D}(\d v+\beta)\,\Big[\d u+\omega + W(\d v+\beta)\Big]+e^{2 G}\d s^2_4+e^{2 \hat{G} }\d \hat{s}^2_{4}\, .
\label{metric244}
\ee
Eventually we will take  $\hat{Y} = T^4$ with a flat metric, but the following arguments work also when  $\hat{Y}$ has  Ricci-flat hyper-K\"ahler metric $\d s^2_{\rm K3}$.  

On the other hand, being $\d s^2_4$ a priory completely arbitrary, the warping $e^{2 G}$ is a `pure-gauge' degree of freedom, in the sense that 
we have the freedom to perform the  gauge transformation
\be\label{warpgauge}
G\rightarrow \Lambda\quad~~\text{and}\quad~~~~~~\d s^2_4\rightarrow e^{-2\Lambda}\d s^2_4\,,
\ee 
with $\Lambda$ an arbitrary function. We will  fix this redundancy by making a convenient gauge choice suggested by the equations.

\vspace{0.2cm}

At different steps in our derivation, we will also make the simplifying assumption that the backgrounds are `isotropic' along $\hat{Y}$.
Configurations describing more general states of the D1-D5-P system can in principle be included by generalizing our results.

\subsection{Restricted spinorial structure}
\label{projections}

We now derive the restrictions that a D1-D5-P (marginal) bound state puts on the form of the Killing spinors. 
In this section we will assume that the NSNS two-form has no legs on $\hat{Y}$ and
\be
\label{isomB}
\mathcal{L}_K B = 0\, .
\ee
According to the structure of the metric \eqref{metric244},  we take a factorised form for the eight-dimensional gamma-matrices\footnote{From now on we use $\gamma^i$ and $\hat\gamma^a$ for the 4D gamma matrices in $Y$ and $\hat{Y}$ respectively and introduce a subscript when we refer to the 8D gamma matrices.}
\be
\label{gamma4+4}
\gamma^{\ul{i}}_{(8)}=\gamma^{\ul{i}}\otimes \bbone \ ,  
\quad
\gamma^{\ul{4+a}}_{(8)}=\gamma_{(4)}\otimes \hat\gamma^{\ul{a}} \, , 
\ee
and the eight-dimensional spinors $\eta_I$
\beq
\label{spinors4+4}
\eta_I  =\zeta^+_I  \otimes \hat \zeta^+_I  + \zeta_I^{+ {\rm c}} \otimes \hat \zeta_I^{+  {\rm c}} + \zeta^-_I \otimes \hat \zeta^-_I + \zeta_I^{-  {\rm c}} \otimes \hat \zeta_I^{-  {\rm c}} \,  , 
\eeq
with  $\zeta^\pm_\alpha$ and $\hat{\zeta}^\pm_\alpha$ chiral spinors on the four non-compact spatial directions $Y$  and  $\hat{Y}$. The suffix ${}^{\rm c}$ denotes the
conjugation $\zeta^{+  {\rm c}}= C_{(4)} \zeta^{+ *} $ (see Appendix  \ref{app:conv} for notations and conventions). \\

Let us first consider what  are the constraints  a momentum in  the direction of $S^1$ sets on the spinors $\epsilon_I$.
A wave propagating left-wise in the direction $y$ is supersymmetric if 
\beq
\label{mompr}
 \Gamma^{\ul {u v}} \epsilon_I = \epsilon_I \, . \\
\eeq
This corresponds to the existence of the null Killing vector $K$ and is automatically satisfied by the spinor ansatz \eqref{ansatz1}
\be
\label{ansatz1bis}
\epsilon_I=\left(\begin{array}{c} 1 \\ 0 \end{array}\right)\otimes \eta_I   \qquad \qquad (I = 1,2) \, . 
\ee

\vspace{0.2cm}

We then require that a D1-brane probe, filling the $(u,v)$ directions and sitting at a generic point of the internal eight-dimensional space, is always supersymmetric. 
The supersymmetry condition for a probe D1-brane is 
\be\label{susyD1}
\Gamma_{\rm D1}\epsilon_2= - \epsilon_1 \, . 
\ee 
In general,
\be
\Gamma_{\rm D1}=\frac{\epsilon^{\alpha\beta}}{2 \sqrt{-\det\iota^*(g-B)}}\,(\Gamma_{\alpha\beta}- B_{\alpha\beta}) \, , 
\ee
where  $\sigma^\alpha$, with  $\alpha=0,1$,  denote  world-volume coordinates,  $\Gamma_\alpha=\partial_\alpha X^M E_M^{\ul M}\Gamma_{\ul M}$ and $\iota^* g\equiv g_{\alpha\beta}\d\sigma^\alpha\d\sigma^\beta$ and
 $\iota^* B\equiv \frac12 B_{\alpha\beta}\d\sigma^\alpha\wedge \d\sigma^\beta$ are the  pull-back of the metric and NSNS two-form on the world-volume.  Notice that, since we consider D1-brane probes with no induced charges,
the world-volume gauge field is zero.

We can use the supersymmetry condition \eqref{susysum1a} and the fact that $K$ is a symmetry of the solution to determine the form of the NSNS two-form.  Indeed (\ref{susysum1a})
is solved by taking
\be
\iota_KB=-\chi=-e^{2D}\cos2\theta\, (\d v+\beta) \, , 
\ee
which means
\be
\label{NS2f}
B= -e^{2D}\cos2\theta\,(\d u+\omega) \wedge(\d v+\beta)+ b \wedge  (\d v+\beta) + \calb \,,
\ee
where $b$ and $\calb$ are a 1- and 2-form on the internal eight-dimensional space $X$.

By using the metric \eqref{metric} and \eqref{NS2f} we find that, in our case,
\be
\Gamma_{\rm D1}=\frac{1}{\sin 2 \theta} (\Gamma_{\ul{u v}}+ \cos 2 \theta \,  \bbone_{(10)}  )= \frac{1}{\sin 2 \theta} (- \sigma_3+  \cos 2 \theta \,  \bbone_{(2)}  )\otimes \bbone_{(8)} \, , 
\ee
where in the last step we have used the $2+8$ decomposition  \eqref{gamma2+8} of the gamma matrices. Hence, the projection (\ref{susyD1}) on the spinors (\ref{ansatz1bis}) reduces to
\be
\label{pr2sp}
 \frac{\sin \theta}{\cos \theta} \,\eta_2=\eta_1 \, ,
\ee
which implies that the  two internal eight-dimensional spinors $\eta_{1,2}$ are proportional.  Comparison with the normalisation condition \eqref{normspin}  fixes\footnote{Notice also that Majorana conditions $\eta_\alpha^\dagger=\eta_\alpha^TC_{(8)}$ imply that the proportionality constants must be real.}
\be
\label{ansatz2}
\eta_1=2^{\frac{1}{4}}\sin\theta \,  \eta \quad~~~\,,\quad ~~~ \eta_2=2^{\frac{1}{4}}\cos\theta \, \eta \, , 
\ee
where the Majorana-Weyl  spinor  $\eta$ has positive chirality and unitary norm.

\vspace{0.2cm} 

Finally we have  to impose that  a D5-brane probe wrapping  $\hat Y$, extending along the $(u,v)$ direction and sitting at any point of  $\hat{Y}$ is supersymmetric
\be
\label{D5pr}
\Gamma_{\rm D5}\epsilon_2=\epsilon_1 \, .
\ee
As for the D1-brane, there is no purely world-volume gauge field.  Since we are assuming vanishing $B$-field along $\hat{Y}$,
we can rewrite the D5 projector in terms of the D1 one
\be
\Gamma_{\rm D5}=\Gamma_{\rm D1}(\bbone\otimes\bbone\otimes\hat\gamma_{(4)}) \, .
\ee
We see that \eqref{D5pr}  reduces the spinor  \eqref{spinors4+4} to  
\be
\label{ansatz22}
\eta=\zeta\otimes\hat\zeta+ \zeta^{\rm c} \otimes \hat\zeta^{\rm c} \, ,
\ee
where $\zeta$ and $\hat\zeta$ have positive chirality in four dimensions  ($\gamma_{(4)}\zeta=\zeta$, $\hat\gamma_{(4)}\hat\zeta=\hat\zeta$) and we can choose them to have unitary norm.

\bigskip

With the ansatz (\ref{ansatz2})-(\ref{ansatz22}) for the eight-dimensional spinors, the polyform $\Psi$ takes the form (\ref{splitpsi}) with
\be
\Phi =  \frac{1}{\sqrt{2}}(1+\Omega+e^{4G+4\hat  G}{\rm vol}_4 \wedge \hat{{\rm vol}}_{4} )\,,
\ee
where $\Omega$ is the four-form defining the Spin(7) structure associated with $\eta$  (see Appendix \ref{app:spin7} for some definitions and properties of Spin(7) structures)
\be\label{Omegared}
\Omega =  e^{4 G}{\rm vol}_4+e^{4 \hat G} \hat{{\rm vol}}_{4}-e^{2G+2 \hat G}\sum^3_{A=1}J_A\wedge \hat J_A \, .
\ee
Furthermore, the forms (\ref{Omega4f}) can be expressed in terms of $\Omega$ too
\be
\Omega^{(1)}=\sqrt{2}\sin^2\theta\,\Omega\quad~~~\,,\quad~~~\Omega^{(2)}=\sqrt{2}\cos^2\theta\,\Omega\, .
\ee

In the equation  above ${\rm vol}_4$ and $ \hat{{\rm vol}}_4$ denote the volume forms on  $Y$ and $\hat{Y}$ respectively, while $J_A$ and
 $\hat J_A$ ( $A=1,2,3$) are two triplets of  two-forms  which define two SU(2) structures on $Y$ and $\hat Y$, respectively.  They are built as bilinears of the spinors  $\zeta$ and $\hat\zeta$ as follows
 \begin{align}
 \label{hyperkahlerstr}
 J_1+\ii J_2 &= \frac12\,\zeta^TC_{(4)}\gamma_{ij}\zeta \, \d x^i \wedge \d x^j~,  \qquad 
 J_3 = \frac\ii2\,\zeta^\dagger\gamma_{ij}\zeta\,  \d x^i \wedge \d x^j  \, ,\\
 -\hat{J}_1+\ii \hat{J}_2&=\frac12\, \hat\zeta^T\hat C_{(4)}\hat\gamma_{ab}\hat\zeta\, \d \hat x^a \wedge \d \hat x^a\,, \qquad \hat J_3 = \frac\ii2\,\hat\zeta^\dagger\hat\gamma_{ab}\hat\zeta\,  \d \hat x^a \wedge \d \hat x^a\, .
  \end{align}
  
 Both triplets of two-forms are anti-self dual with respect to the corresponding metrics
\be\label{asdj}
*_4 J_A=-J_A \qquad  \mbox{and} \qquad  \hat*_4\hat J_A=-\hat J_A \qquad  A=1,2,3\,,
\ee
and satisfy the usual properties of almost hyperk\"ahler structures, e.g.
\be\label{quateq}
(J_A)^i{}_k(J_B)^k{}_j=\sum_B\epsilon_{ABC}(J_C)^i{}_j-\delta_{AB}\delta^i_j\,,
\ee
or equivalently
\be
J_A\wedge J_B = -2\,\delta_{AB}\,\mathrm{vol}_4\,.
\ee
The same equations hold for $\hat{J}_A$.

\subsection{Minimal set of equations for our  ansatz}
\label{genans}

The second part of the paper will be devoted to the construction of new examples of D1-D5-P geometries describing microstates of a black hole. In this section
we specify the general supersymmetry conditions of Section \ref{summary} to the restricted spinorial structure we worked out in 
Section \ref{projections} and we  identify a minimal set of equations which need to be solved in order to get a full supergravity solution. 

\vspace{0.2cm}

For concreteness, in the rest of the paper we will always take $\hat{Y}= T^4$, choosing $\d \hat s^2_4$ to be a flat metric on it\footnote{We stress once again that, with the two assumptions recalled after~\eqref{3.26}, everything we will say is actually valid for $\hat Y=$K3 and Ricci-flat $\d \hat s^2_4\equiv \d s^2_{\rm K3}$ as well.}.  Then we can take  
the triplet $\hat J_A$ to be closed
\be\label{3.26}
\d \hat J_A=0 \, , 
\ee
so that $T^4$ is endowed with  a hyper-k\"ahler structure. 
We will also assume that 
all the fields depend just on the $(u,v,x^i)$ coordinates along the six-dimensional space $\mathbb{R}^{1,1}\times Y$. 

Finally we restrict our analysis  to backgrounds that  are `isotropic' along $T^4$. In other words, we impose that $H$ has legs just along   $\mathbb{R}^{1,1}\times Y$ and 
that the RR-flux polyform $F$  splits as follows
\be\label{Ftotan2}
F_{\rm tot}=F+\tilde F\wedge \hat{\rm vol}_4 \, , 
\ee
where 
\begin{subequations}\label{6Fexp}
\begin{align}
F&\equiv F_1+F_3+F_5\label{6Fexpa} \, , \\
\tilde F&\equiv \tilde F_1+\tilde F_3+\tilde F_5 \label{6Fexpb}
\end{align}
\end{subequations} 
 have legs along $\mathbb{R}^{1,1}\times Y$ only, and  $\hat{\rm vol}_4$  is the volume form associated with the flat metric $\d \hat s^2_4$. To avoid confusion, in this section,
we denote the  ten-dimensional RR fluxes by the subscript $_{\rm tot}$.
The ten-dimensional self-duality condition  $ * \lambda(F_{\rm tot}) = F_{\rm tot}$  translates into the  six-dimensional Hodge-duality
\be\label{6SD2}
\tilde F=e^{4\hat G} *_6 \lambda(F) \, ,
\ee
where $*_6$ uses the complete, warped, six-dimensional metric. Notice that, by isotropy, $F$ and $\tilde F$ must satisfy the following six-dimensional Bianchi identities/equations of
motion
\begin{subequations}\label{6BI0}
\begin{align}
\d_H F&=0 \, , \label{6BI}\\
\d_H\tilde F&=0 \, , \label{6BI2}
\end{align}
\end{subequations}
where we recall that $\d_H\equiv \d-H$. 

\bigskip

The BPS equations of  Section \eqref{summary} can be used to derive a general ansatz (provided the assumptions we made before) describing the D1-D5-P geometries we want to study.  In particular the metric and fluxes can be expressed in  terms of a reduced number of independent fields satisfying a simplified set of equations.  The detailed derivation
of such a minimal ansatz can be found in Appendix \ref{app:genansatz}. Here we simply discuss the main steps of that derivation and the final results,  omitting several technical details.

\bigskip

Let us consider first  \eqref{susysum1a}. 
As already discussed in Section \eqref{projections},  this can be used to derive the most general for of the NSNS two-form, under the assumption that $\mathcal{L}_K B = 0$.
\be
\label{Bsolt}
B= -e^{2D}\cos2\theta\,(\d u+\omega) \wedge(\d v+\beta)+ b \wedge  (\d v+\beta)+ \calb \,,
\ee
where $b=b_i\d x^i$ and $\calb =\frac12 \calb_{ij}\d x^i\wedge \d x^j$ have legs just along $Y$. 

\vspace{0.2cm} 

Similarly the `isotropic' components of \eqref{susysum1b} can be used to determine the RR potentials.  We  define 
\be
F=\d_H C  \qquad \,,\qquad \tilde F=\d_H\tilde C \, ,
\ee  
where $C=C_0+C_2+C_4$ and  $\tilde C= \tilde C_0+\tilde C_2+\tilde C_4$  are  $u$-independent. Then the most general solution of \eqref{susysum1b} for the potentials is
\begin{subequations}
\label{RRpotsolt}
\begin{align}
C =&-e^{2D}(\d u+\omega)\wedge (\d v+\beta)\wedge (e^{-\phi}\sin2\theta+\cos2\theta\, \calc)+ c \wedge (\d v+\beta) +\calc\,, \\
\tilde C =&-e^{2D}(\d u+\omega)\wedge (\d v+\beta)\wedge \big(e^{4\hat G-\phi}\sin2\theta+\cos2\theta\, \tilde\calc\big)+  \tilde c \wedge (\d v+\beta ) +\tilde \calc \,,
\end{align}
\end{subequations}
where we have introduced the following polyforms on $Y$
\be
\begin{aligned}
c &\equiv  c_1+ c_3  \, ,\cr  
\tilde c&\equiv\tilde c_1+\tilde c_3 \, , \cr  
\calc&\equiv  \calc_0+  \calc_2+  \calc_4  \, , \cr
\tilde \calc&\equiv\tilde \calc_0+\tilde \calc_2+   \tilde \calc_4 \, .
\end{aligned}
\ee

In order to proceed, we observe that the gauge freedom (\ref{warpgauge}) can be conveniently fixed by imposing that 
\be\label{Gfix}
e^{-2G}=e^{2D+2\hat G-\phi}\sin2\theta \, .
\ee
Then, the isotropy condition tells us the that the `non-isotropic' components of \eqref{susysum1b}  must  vanish. This gives conditions involving the two-forms $J_A$ 
\be
\label{iscondt}
\cald J_A-\dot\beta\wedge J_A=0
\ee
and the one-form $\beta$
\be
*_4\cald\beta=\cald\beta \, .
\label{betasd}
\ee
The first tells  us that the non-trivial $v$-dependence of the background constitutes a potential obstruction to the integrability of the almost hyperk\"ahler structure on $Y$.
The second conditions is simply the self-duality of  $\cald\beta$.

By using  \eqref{susysum2a} and  \eqref{susysum2b} one can show that the dilaton is given by
\be
\label{phiandD}
e^\phi = \frac{e^{2 \hat G}}{\sin2\theta} \, , 
\ee
up to an arbitrary overall constant factor, which we have chosen to be 1 for simplicity. 

It is convenient to explicitly solve the two relations (\ref{Gfix}) and (\ref{phiandD}) by expressing  the three warpings $e^{2D}$, $e^{2G}$ and $e^{2\hat G}$, the  angle $\theta$ and the dilaton $e^\phi$ in terms of three independent functions $Z, \tilde Z$ and $Z_b$\footnote{It can be useful to list the inverse relations too:
$Z=e^{2G+2\hat G}$, $\tilde Z=e^{2G-2\hat G}$ and $Z_b=\cos2\theta e^{2G}$, with $e^{2G}$ given by (\ref{Gfix}).}:
\be\label{qarp}
\begin{aligned}
 &e^{2D}=\frac{\alpha}{\sqrt{Z \tilde Z}}\,,\quad~~~~~e^{2G}=\sqrt{Z \tilde Z}\,,\quad~~~~~~~ e^{2\hat G}=\sqrt{\frac{Z}{\tilde Z}}\,,\\
 &\cos2\theta=\frac{Z_b}{\sqrt{Z \tilde Z}}\,,\quad~~~~~~~~~~ e^{2\phi} = \alpha\,\frac{Z}{\tilde Z}\,,
 \end{aligned}
\ee
where we have introduced a further function
\be
\alpha\equiv \frac{Z \tilde Z}{Z \tilde Z-Z^2_b}=\frac{1}{\sin^22\theta}\,.
\ee
By combining  \eqref{susysum2a} and  \eqref{susysum2b} with the self-duality condition for the RR field-strengths  we can also fix
\be
\label{Cscalarst}
\calc_0 = \frac{Z_b}{Z}  \quad~~~\,,\quad~~~ \tilde\calc_0= \frac{Z_b}{\tilde Z}\,,
\ee
up to an additive constant, which we set equal to zero.
We then see that part of the supersymmetry conditions can be used to fix the form of all scalars in our ansatz in terms of just the three  functions $Z$, $\tilde Z$ and $Z_b$.  
\vspace{0.2cm} 

In order to write the remaining supersymmetry conditions in a more inspiring form, let us introduce the anti-self-dual two form~\cite{Gutowski:2003rg}
\be
\psi\equiv  \frac{1}{8}\,\epsilon^{ABC}(J_A)^{i j}  ( \dot{J}_B)_{ij}  J_C \, ,
\ee
which measures the rotation of the triplet $\{J_A\}$ under the $v$-flow, and a set of three two-forms
\be\label{thetadef2}
\begin{aligned}
& \Theta = \dot{\calc}_2 + \cald c_1-\dot{\beta}\wedge  c_1 \, , \\ 
& \tilde \Theta =\dot{\tilde\calc}_2 + \cald\tilde c_1- \dot{\beta}\wedge \tilde c_1 \, , \\
& \Theta_b = \dot{\calb} + \cald b -  \dot{\beta}\wedge b\,.
\end{aligned}
\ee
Then, the self-duality of the RR fields  implies that the anti-self-dual components of $\Theta$, $\tilde \Theta$ and $\Theta_b$ are proportional  to  $\psi$ 
\begin{subequations}\label{thetaASD}
\begin{align}
& (1-*_4)\Theta =2\,\tilde Z\,\psi\label{thetaASDa}\, , \\
& (1-*_4)\tilde \Theta=2\,Z\,\psi\label{thetaASDb}\, , \\
& (1-*_4)\Theta_b=2\,Z_b\,\psi\label{thetaASDc}\,,
\end{align}
\end{subequations}
and determines the self-dual component of $\cald\omega$ 
\be\label{omegaeq}
 \cald \omega + *_4  \cald\omega =Z *_4\!\Theta+\tilde  Z\, \tilde \Theta -Z_b\, (\Theta_b +*_4  \Theta_b)- 2\, W \,\cald\beta \ .
\ee
Notice that the self-duality of the r.h.s. of (\ref{omegaeq}) is guaranteed by (\ref{thetaASDa}) and (\ref{thetaASDb}).

\vspace{0.2cm}

The remaining conditions encoded in the self-duality of the RR fields can be shown to reduce to other two sets of equations. The first set is
 \begin{subequations}\label{2formeq}
\begin{align}
& \cald\calc_2-\cald\beta\wedge c_1=*_4(\cald \tilde Z+\tilde Z\dot\beta)\,,\\
& \cald\tilde\calc_2-\cald\beta\wedge \tilde c_1=*_4(\cald Z +Z\dot\beta)\,,\\
& \cald\calb-\cald\beta\wedge b =*_4(\cald Z_b+Z_b\dot\beta)\,.
\end{align}
\end{subequations}
and the  second set is
 \begin{subequations}\label{eqomegat}
\begin{align}
\dot \calc_4+\cald c_3-\dot\beta\wedge c_3-\Theta_b \wedge \calc_2  -(\cald \calb -  \cald\beta\wedge b)\wedge c_1 =&\tilde Z^2\,\frac{\d}{\d v} \left(\frac{Z_b}{\tilde Z}\right){\rm vol}_{4}   \,, \\
\dot{\tilde\calc}_4+\cald \tilde c_3-\dot\beta\wedge \tilde c_3-\Theta_b \wedge \tilde \calc_2  -(\cald \calb -  \cald\beta\wedge b)\wedge \tilde c_1  =& Z^2\,\frac{\d}{\d v} \left(\frac{Z_b}{Z}\right) {\rm vol}_{4}    \,. 
\end{align}
\end{subequations}
In fact, this latter set of equations does not give any additional constraint since it can always be solved locally: one can always choose a gauge in which $\calc_4\equiv \tilde\calc_4=0$ and equations (\ref{eqomegat}) can always be (locally) integrated 
to give the (local) expression for $c_3$ and $\tilde c_3$. 

\bigskip 

To conclude this section it might be useful to summarise what we obtain for  the different gauge invariant fields once we have  implemented the above constraints.
The metric ansatz (\ref{metric244}) can be rewritten as
\be\label{metricfinal}
\d s^2_{(10)} = -\frac{2\alpha}{\sqrt{Z \tilde Z}}\,(\d v+\beta)\,\Big[\d u+\omega + W(\d v+\beta)\Big]+\sqrt{Z \tilde Z}\,\d s^2_4+\sqrt{\frac{Z}{\tilde Z}}\,\d \hat{s}^2_{4}\, .
\ee 
The dilaton is given by
\be\label{dilatonfinal}
e^{2\phi}=\alpha\,\frac{Z}{\tilde Z}\, .
\ee
The NSNS field-strength is
\be\label{Hfinal}
\begin{aligned}
H=&-(\d u+\omega)\wedge(\d v+\beta)\wedge\Big[\cald\Big(\frac{\alpha Z_b}{Z \tilde Z}\Big)-\frac{\alpha Z_b}{Z \tilde Z}\,\dot\beta\Big]\\
&+(\d v+\beta)\wedge\Big(\Theta_b-\frac{\alpha Z_b}{Z \tilde Z}\,\cald\omega\Big)+\frac{\alpha Z_b}{Z \tilde Z}\,(\d u+\beta)\wedge\cald\beta+*_4(\cald Z_b+Z_b\,\dot\beta)\, .
\end{aligned}
\ee
The  RR field-strengths (\ref{6Fexpa}) are
\begin{subequations}\label{Ffinal}
\begin{align}
F_{1}=&\cald\left(\frac{Z_b}{Z}\right)+(\d v+\beta)\wedge\frac{\d}{\d v} \left(\frac{Z_b}{Z}\right)\,,\\
F_{3}=&  -(\d u+\omega)\wedge (\d v+\beta)\wedge\Big[\cald\Big(\frac{1}{Z}\Big)-\frac{1}{Z}\dot\beta+\frac{\alpha Z_b}{Z \tilde Z}\cald\left(\frac{Z_b}{Z}\right)\Big]    \cr
&+(\d v+\beta)\wedge\Big(\Theta-\frac{Z_b}{Z}\Theta_b-\frac{1}{Z}\cald\omega\Big)+\frac{1}{Z}\,(\d u+\omega)\wedge \cald\beta\cr
&+*_4(\cald \tilde Z+\tilde Z\dot\beta)-\frac{Z_b}{Z}*_4(\cald Z_b+Z_b\dot\beta)\,,\\
F_5=&-\frac{\alpha}{Z}(\d u+\omega)\wedge(\d v+\beta)\wedge *_4\Big[\frac{Z_b}{\tilde Z}(\cald \tilde Z+\tilde Z\dot\beta)-\cald Z_b-Z_b\dot\beta\Big]\\
&+\tilde Z^2\frac{\d}{\d v}\Big(\frac{Z_b}{\tilde Z}\Big)(\d v+\beta)\wedge{\rm vol}_{4}\,,
\end{align}
\end{subequations}
while the RR field-strength  (\ref{6Fexpb}) are given by
\begin{subequations}\label{tildeFfinal}
\begin{align}
\tilde F_{1}=&\cald\left(\frac{Z_b}{\tilde Z}\right)+(\d v+\beta)\wedge\frac{\d}{\d v} \left(\frac{Z_b}{\tilde Z}\right)\,,\\
\tilde F_{3}=&  -(\d u+\omega)\wedge (\d v+\beta)\wedge\Big[\cald\Big(\frac{1}{\tilde Z}\Big)-\frac{1}{\tilde Z}\dot\beta+\frac{\alpha Z_b}{Z\tilde Z}\cald\left(\frac{Z_b}{\tilde Z}\right)\Big]    \cr
&+(\d v+\beta)\wedge\Big(\tilde \Theta-\frac{Z_b}{\tilde Z}\Theta_b-\frac{1}{\tilde Z}\cald\omega\Big)+\frac{1}{\tilde Z}\,(\d u+\omega)\wedge \cald\beta\cr
&+*_4(\cald Z+Z\dot\beta)-\frac{Z_b}{\tilde Z}*_4(\cald Z_b+Z_b\dot\beta)\,,\\
\tilde F_5=&-\frac{\alpha}{\tilde Z}(\d u+\omega)\wedge(\d v+\beta)\wedge *_4\Big[\frac{Z_b}{Z}(\cald Z+Z\dot\beta)-\cald Z_b-Z_b\dot\beta\Big]\\
&+Z^2\frac{\d}{\d v}\Big(\frac{Z_b}{Z}\Big)(\d v+\beta)\wedge{\rm vol}_{4}\,.
\end{align}
\end{subequations}
Our general supersymmetric ansatz is completely specified in terms of the fields $\d s^2_4$, $\omega$, $\beta$, $W$, $Z$, $\tilde Z$, $Z_b$, $\Theta$, $\tilde \Theta$ and $\Theta_b$. They have to satisfy (\ref{iscondt}), (\ref{betasd}), (\ref{thetaASD}), (\ref{omegaeq}) and (\ref{2formeq}), which ensure  the Bianchi identities  for the NSNS and RR fields, as well
as the RR self-duality condition (\ref{6SD2})\footnote{In particular, we see we do not need to use (\ref{eqomegat}), which correspond to trivial Bianchi identities.}.  Notice that the definitions (\ref{thetadef2}) and the conditions (\ref{2formeq}), which explicitly involve some (locally defined) RR and NSNS potentials, can be substituted with the conditions
\begin{subequations}\label{thetaglob}
\begin{align}
\cald\Theta-\dot\beta\wedge \Theta=&\frac{\d}{\d v}*_4(\cald \tilde Z+\tilde Z\dot\beta)\,,\\
\cald\tilde \Theta-\dot\beta\wedge \tilde \Theta=&\frac{\d}{\d v}*_4(\cald Z+Z\dot\beta)\,,\\
\cald\Theta_b-\dot\beta\wedge \Theta_b=&\frac{\d}{\d v}*_4(\cald Z_b+Z_b\dot\beta)\, ,
\end{align}
\end{subequations} 
and
\begin{subequations}\label{zetaglob}
\begin{align}
\cald*_4(\cald Z+\dot\beta Z)=& -\tilde \Theta\wedge \cald\beta\,,\\
\cald*_4(\cald \tilde Z+\dot\beta \tilde Z)=& -\Theta\wedge \cald\beta\,,\\
\cald*_4(\cald Z_b+\dot\beta Z_b)=& -\Theta_b\wedge \cald\beta\,,
\end{align}
\end{subequations} 
respectively. Indeed,  (\ref{thetadef2}) and  (\ref{2formeq}) can be regarded as explicit local solutions of the equations (\ref{thetaglob}) and (\ref{zetaglob}). 

\bigskip

As explained in Section \ref{intnull}, in order to obtain a  supersymmetry solution, one needs to further impose the $\ul{vv}$ component of the Einstein equation (\ref{sumEins}). 
By using the parametrization introduced in this section and some of the above constraints imposed by supersymmetry, this reduces to the following equation
\be\label{evvfin}
\begin{aligned}
&*_4\cald *_4 L+2\,\dot\beta_i\, L^i  +\frac14 \frac{Z \tilde Z}{\alpha} \,\dot{g}^{ij} \dot{g}_{ij}-\frac12 \frac{\d}{\d v}\Bigl[ \frac{Z \tilde Z}{\alpha}\,g^{ij} \dot{g}_{ij}\Bigr] \\
&-\dot{Z} \dot{\tilde Z}-Z \ddot{\tilde Z} - \ddot{Z} {\tilde Z}+(\dot{Z}_b)^2 + 2 \,Z_b \ddot{\tilde Z}_b\\
&+\frac{1}{2} *_4 \Big[(\Theta-\tilde Z\psi)\wedge (\tilde \Theta-Z \psi) - (\Theta_b-Z_b\psi)\wedge (\Theta_b-Z_b\psi)\\
&\qquad + \frac{Z \tilde Z}{\alpha}\, \psi\wedge \psi-2 \psi\wedge \cald\omega\Bigr]=0\,,
\end{aligned}
\ee
where we recall that
\be\label{defL}
L = \dot{\omega} + W\,\dot{\beta}-\cald W\,,
\ee
$g_{ij}$ are the components of the metric on $Y$: $\d s^2_4 = g_{ij}\,\d x^i\d x^j$, and $\dot{g}^{ij}\equiv \frac{\d}{\d v} g^{ij}$.

\bigskip

It is useful to examine some limits of the general ansatz we have found above to clarify how previously known solutions embed in it. When the metric and the gauge fields are taken to be independent of $v$, the solution reduces to the one of \cite{Giusto:2012gt, Vasilakis:2012zg} and it can be reduced to $\mathcal{N}=2$ 5D supergravity coupled to three vector multiplets. When $Z_b=b=\mathcal{B}=0$ one re-obtains the ansatz of \cite{Cariglia:2004kk,Bena:2011dd}, which is equivalent to  6D supergravity coupled to an anti-self-dual tensor multiplet (if one further restricts $Z=\tilde Z$, $c_1=\tilde c_1$, $\calc_2=\tilde\calc_2$, one reduces to minimal 6D supergravity, studied in \cite{Gutowski:2003rg}). 

We would like to conclude this section by emphasizing  that, in order to find explicit solutions, the minimal set of equations we listed above can 
 be conveniently organized in a way which highlights a hidden linear structure \cite{Bena:2011dd}.  This is discussed in Appendix \ref{app:summary}.

\subsection{Supersymmetry of the solution}

The supergravity backgrounds described  in Section \ref{genans} are 1/8 supersymmetric, that is they preserve four supercharges,  both for  $\hat Y = T^4$ or K3. In order to understand this point we need to count the number of independent degrees of freedom of the spinors $\epsilon_1$ and $\epsilon_2$ satisfying the supersymmetry conditions.
According to our ansatz (\ref{ansatz1bis}) and (\ref{ansatz2})-(\ref{ansatz22}) these are given by
\be
\label{ansatzbis}
\epsilon_I = N_I \begin{pmatrix} 1 \\ 0 \end{pmatrix} \otimes ( \zeta \otimes\hat\zeta +\zeta^{\rm c}\otimes\hat\zeta^{\rm c} )\,,
\ee
where the spinors $\zeta$ and $\hat{\zeta}$ are positive chirality  spinors  of unitary norm on $Y$ and $T^4$,  respectively. The functions $N_I$ are given
in \eqref{ansatz2}:  $N_1 =  2^{1/4} \sin \theta$ and $N_2 =  2^{1/4} \cos \theta$. 

\vspace{0.2cm}

In our approach we trade spinors for forms built as bilinears in the spinors.  Indeed, the complete information on the spinorial ansatz (\ref{ansatzbis}) is carried by the angle $\theta$ and the four-form $\Omega$ introduced in (\ref{Omegared}). In turn,
the information on $\zeta$ and $\hat{\zeta}$ is encoded in the two almost hyperk\"ahler structures $(J_1,J_2,J_3)$ and $(\hat J_1,\hat J_2,\hat J_3)$  (see \eqref{hyperkahlerstr}),
which enter $\Omega$ through the combination
\be
\label{diagJ}
\sum_A J_A\wedge\hat J_A \, .
\ee

Now, it is immediate to check that all the equations for our backgrounds are invariant under separate rigid  SU(2)$\simeq$ SO(3) rotations  of $J_A$ and $\hat J_A$, say ${\rm SU}(2) \times \hat{\rm SU}(2)$. Such transformations do not change the metric and the other bosonic fields but do generically transform the form $\Omega$, which is  left invariant only under by the diagonal subgroup ${\rm SU}(2)_{\text{diag}} \subset {\rm SU}(2) \times \hat{\rm SU}(2)$ which preserves the combination (\ref{diagJ})\footnote{The SU(2) rotations translate into analogous rotations of the spinors $\zeta$ and $\hat\zeta$, and again it is easy to see that  \eqref{ansatzbis} is left invariant only by the diagonal subgroup  ${\rm SU}(2)_{\text{diag}}$.}.
This means that these rigid ${\rm SU}(2) \times \hat{\rm SU}(2)$ transformations produce a three-parameters family of inequivalent normalized Killing spinors, locally identifiable  with the coset  $[{\rm SU}(2) \times \hat{\rm SU}(2)]/{\rm SU}(2)_{\text{diag}}$.  We also have the freedom of
a constant rescaling of the Killing spinors.  If we take into account this extra parameter,  we obtain a four-parameter family of Killing spinors, corresponding to four supercharges preserved  by our backgrounds.

\section{An exact solution}
\label{solutionexample}

The supergravity analysis of the previous sections provides a general
framework for the construction of supersymmetric type IIB solutions with a null isometry.
In particular it can be specialised  to the study of type IIB backgrounds containing a $T^4$ and isotropic in  $T^4$. The main physical
application that motivated this analysis is the construction of the
generic solution carrying D1, D5 and P charges. As mentioned in the
Introduction, it would be very interesting to see whether the subset
of D1-D5-P microstates that are isotropic in $T^4$
is sufficient to account for a finite fraction of the total entropy
associated with the given asymptotic charges.

In this Section we present  the first example of a configuration of this kind: an exact,  completely regular, horizonless,   $v$-dependent solution of the supergravity equations of Section \ref{summary}
carrying D1, D5 and P charges, with a non-trivial profile for  all type IIB fields. It also
admits an AdS limit which is dual to a known state of the D1-D5
CFT. 
Our main  purpose  is to show that it is possible
to find such explicit $v$-dependent solutions\footnote{The example
  discussed in~\cite{Ford:2006yb} falls in the restricted ansatz
  of~\cite{Bena:2011dd} and a full analysis of the regularity
  conditions of this case has still to be performed. Another very
  intesting family of a $v$-dependent geometries has been recently
  discussed in~\cite{Mathur:2012tj,Lunin:2012gp}; these solutions are
  somehow complementary to those discussed in this paper as they are
  non-trivial along  $T^4$ and are identical to the two-charge
  configurations for the 6D part. A class of $v$-dependent but unbound solutions has been found in~\cite{Niehoff:2012wu,Vasilakis:2013tjs}.}, rather than to discuss how to build them.  That is why we will  give  the
explicit form of the solution but only briefly sketch the method that
was used to construct it (this is basically the approach
of~\cite{Mathur:2011gz}, just implemented at the full non-linear
level). A more detailed presentation of the solution generating
technique, of the regularity analysis and of the duality with the
D1-D5 CFT, together with possible generalizations, will be presented
in a forthcoming work.

The solution we obtain can be seen as the non-linear completion of an
approximate solution derived by Mathur, Saxena and Srivastava (MSS) in~\cite{Mathur:2003hj}, which
represented the first example of a microstate geometry for the three-charge black hole. We will review below the approximate construction
of~\cite{Mathur:2003hj}, and then outline our method to promote that
solution to an exact one of the non-linear supergravity
equations.

Note:  To make contact with the existing literature on D1-D5-P microstates, in this section we will use a different notation than in the previous part of the paper.
The switch of notation and the supergravity equations in the new variables are give in  Appendix~\ref{app:summary}. From now on we will refer to equations in that appendix.

\subsection{The perturbative solutions of MSS}

The three-charge solution of MSS~\cite{Mathur:2003hj} is realized as a
perturbation around a particular two-charge background. Thus we begin
with a very brief summary of the D1-D5 1/4-BPS microstates. These
geometries were derived starting from the solution describing the
back-reaction of a vibrating  F1-string~\cite{Dabholkar:1995nc, Callan:1995hn}. 
When the asymptotic  10D geometry is 
$\mathbb{R}^{1,4}\times S^1\times T^4$ it is possible to perform a U-duality transformation mapping the
F1-P charges into the D1-D5 charges.  One obtains a microstate 
defined by a curve $g_i(v')$ in $\mathbb{R}^4$. We are interested in the
Lunin-Mathur solution~\cite{Lunin:2001fv, Lunin:2001jy} which is
defined by the circular profile
\be
\label{circle}
g_1(v') = a \cos \frac{2\pi\,v'}{L} \,,\quad g_2(v') = a \cos \frac{2\pi\,v'}{L} \,,\quad g_3(v') = g_4(v')=0\,.
\ee
Here
\be \label{Q1Q5Ra}
a= \frac{\sqrt{Q_1 Q_5}}{R}\,,\quad L = \frac{2\pi\, Q_5}{R}\,,
\ee
where $R$ is the radius of the direction $y$ common to the D1 and D5
branes, and $Q_1$ and $Q_5$ are the charges for $N_1$ D1 and $N_5$ D5
branes. The $v'$ appearing in the equation above has to be thought as
a parameter along the profile $g_i(v')$, and is not to be confused
with the space-time coordinates $v$, used in the supergravity
sections, which is related to the time coordinate $t$ and the $S^1$
coordinate $y$ as in (\ref{timelike}).

In this and the next subsection we will restrict, as
in~\cite{Mathur:2003hj}, to the particular case in which
$Q_1=Q_5=Q$. This ensures that the 6D description of the solution
in~\eqref{4Dbase}--\eqref{omega0} is simpler as it can be described by 
minimal supergravity. In this case, the Lunin-Mathur geometry with
circular profile can be written in terms of the ansatz~\eqref{ansatzsummary} as
follows
\begin{subequations}
\begin{align}
\d s^2_4 &= (r^2+a^2 \cos^2\theta)\Bigl(\frac{\d r^2}{r^2+a^2}+\d \theta^2\Bigr)+(r^2+a^2)\sin^2\theta\, \d \phi^2+r^2 \cos^2\theta\,\d \psi^2\,,\label{4Dbase}\\
\beta&= \frac{R\,a^2}{\sqrt{2}\,(r^2+a^2\cos^2\theta)}\,(\sin^2\theta\,\d\phi - \cos^2\theta\,\d\psi)\,,\label{beta0}\\
Z_1&= Z_2=1+\frac{Q}{r^2+a^2\cos^2\theta}\,,\quad a_1=a_2=0\,,\label{Z12}\\
\omega&=\frac{R\,a^2}{\sqrt{2}\,(r^2+a^2\cos^2\theta)}\,(\sin^2\theta\,\d\phi + \cos^2\theta\,\d\psi)\label{omega0}\,,\quad \mathcal{F}=0\,,\\
Z_4&= a_4=\delta_2=0\,.
\end{align}
\end{subequations}
Note that the 4D base metric $\d s^2_4$ in~(\ref{4Dbase}) is just flat
$\mathbb{R}^4$ written in non-standard coordinates, in which the 10D
metric takes its simplest form.

\bigskip

For $r\sim\sqrt{Q}$ a curved
region of space-time, named the ``throat'', opens up.  Contrary to the
``naive'' extremal black hole geometry (which corresponds to the 
case $a=0$), the throat ends smoothly after a coordinate distance of order
$Q/a$. As usual, the decoupling (or ``near-horizon'') limit
corresponds to the case in which one focuses on the region inside a
throat whose length is much larger than its width. Quantitatively this
approximation requires
\be
r\ll \sqrt{Q}\,,\quad a\ll \sqrt{Q}\,.
\ee
In this limit, the ``1'' in the expression~(\ref{Z12}) for $Z_1$ and
$Z_2$ can be neglected:
\be\label{Z12nh}
Z_1^{\rm nh}=Z_2^{\rm nh}=\frac{Q}{r^2+a^2\cos^2\theta}\,,
\ee
while all the other geometric data are left unchanged. One can see
that the resulting 10D geometry reduces to $AdS_3\times S^3\times T^4$
after the coordinate redefinition
\be
\label{spectralflow}
\phi\to \phi + \frac{t}{R}\,,\quad \psi \to \psi + \frac{y}{R}\,.
\ee

\vspace{0.2cm}

According to the general AdS/CFT paradigm, the full string (or M)
theory in an AdS space arising from a ``near-horizon'' limit should be
dual to the CFT describing the low-energy approximation of the theory
living on the branes used in the construction. In our case, we have to
deal with the 1+1 dimensional CFT with central charge $6 N_1
N_5$~\cite{Maldacena:1996ky,David:2002wn} that captures the low energy dynamics of the open
strings ending on the D1 and D5 branes. Let us recall that this CFT
has an $SU(2)_L\times SU(2)_R$ R-symmetry, corresponding to
rotations of $\mathbb{R}^4$, whose affine generators are $J_n^3$, $J_n^\pm$
and $\bar J_n^3$, $\bar J_n^\pm$. According to the standard AdS/CFT dictionary, the ``empty''
$AdS_3\times S^3\times T^4$ space corresponds to the vacuum state of
the CFT, which is in the NS-NS sector, while the geometry defined by
Eq.~\eqref{circle} is dual to the RR ground state of the D1-D5 CFT
with maximal values of $J_0^3$ and $\bar J_0^3$ (i.e.~the highest
weight state in each $SU(2)$ given the total angular momentum of the
geometry). Then, from the point of view of the CFT, the coordinate
change (\ref{spectralflow}) realizes the spectral flow from the RR
sector to the NSNS sector \cite{Balasubramanian:2000rt}, since it
connects the geometry dual to a RR ground state to the ``empty''
$AdS_3\times S^3\times T^4$, dual to the NSNS ground state.

Non-trivial chiral primaries of the CFT are more easily described by
reducing to six-dimensional supergravity, and are represented by
supergravity perturbations of the $AdS_3\times S^3$ background. This
is the point of view adopted by the authors of~\cite{Mathur:2003hj},
who consider a 6D supergravity comprising the gravity multiplet
(whose bosonic part contains the metric, a self-dual 2-form $C$ and a
vector) and a tensor multiplet (which includes an ``anti-self-dual''
2-form $B$, a scalar $w$ and a vector).  The metric and $C$ describe
the $AdS_3\times S^3$ background:
\begin{subequations}
\begin{align}
\d s^2_6 & =  - \frac{r^2+a^2}{Q}\,\d t^2+\frac{r^2}{Q}\, \d y^2+Q\,\frac{\d r^2}{r^2+a^2}+ Q\, (\d\theta^2 + \cos^2\theta\, \d\psi^2 + \sin^2\theta\, \d\phi^2)\,,\\
C & =-\frac{r^2+a^2}{Q}\,\d t\wedge \d y - Q\,\cos^2\theta\,\d \phi\wedge \d\psi\, .
\end{align}
\end{subequations}
Then, following~\cite{Mathur:2003hj}, we switch on a perturbation that
sits in the tensor multiplet and only excites the fields $B$ and
$w$. The equations satisfied by $B$ and $w$ are the 6D supergravity
equations {\it linearized} around the $AdS_3\times S^3$ background and
are given by
\be
 \d B+*_6 \d B+w\,\d C=0\,,\quad \d*_6 \d w - 2 \d B \wedge \d C=0\,.
\ee
The explicit form of the perturbation\footnote{In all linearised
  solutions we follow~\cite{Mathur:2003hj} and complexify the field
  describing the perturbation: both the real and the imaginary parts
  of this field represent valid solutions. Of course, at the
  non-linear level we will always have to work with real fields.}
is~\cite{Mathur:2003hj}
\begin{subequations}
\begin{align}
\label{pert1}
w &= \frac{c_l}{Q}\,e^{-2\,i \,l\,(\phi+ \frac{t}{R})}\,\frac{\sin^{2l}\theta}{(r^2+a^2)^l}\,,\\\label{pert2}
B &= \frac{c_l}{2}\,e^{-2\,i \,l\,(\phi+ \frac{t}{R})}\,\frac{\sin^{2l}\theta}{(r^2+a^2)^l}\,\Bigl[-\frac{r^2}{Q^2}\, \d t\wedge \d y - \frac{i}{R}\,\frac{r}{r^2+a^2}\, \d r\wedge \d y \nonumber\\
&\qquad- \cos^2\theta \, \d\phi\wedge \d\psi - i \frac{\cos\theta}{\sin\theta}\, \d\theta\wedge \d\psi\Bigr]\,.
\end{align}
\end{subequations}
This perturbation is dual to a chiral primary state
$|\Psi\rangle_{NS}$ identified by the quantum numbers
\be
j_{NS}=h_{NS}=l\,,\quad \bar j_{NS}=\bar h_{NS}=l\,,
\ee
where $j_{NS}$ and $\bar j_{NS}$ are the eigenvalues of $J_0^3$ and
$\bar J_0^3$ and $h_{NS}$ and $\bar h_{NS}$ are the eigenvalues of the
Virasoro generators $L_0$ and $\bar L_0$.

For later convenience, it is also useful to analyze the corresponding solution in the RR sector. $|\Psi\rangle_{NS}$ maps, via the inverse of the spectral flow
transformation
\be
\label{inversespectralflow}
\phi\to \phi - \frac{t}{R}\,,\quad \psi \to \psi - \frac{y}{R}\, , 
\ee
into a RR ground state $|\Psi\rangle_{R}$ with\footnote{Spectral flow
  transforms the CFT quantum numbers as $h_R =
  h_{NS}-j_{NS}+\frac{c}{24}$, $j_R=j_{NS}-\frac{c}{12}$, with $c=6\,
  N_1 N_5$ for the D1-D5 CFT. The terms proportional to $c$ are
  associated with the background and thus, for the perturbation alone,
  one has $h_R = h_{NS}-j_{NS}$, $j_R=j_{NS}$.}
\be
j_R=l\,,\quad h_R =0\,,\quad \bar j_R=l\,,\quad \bar h_R =0\,.
\ee
The geometry corresponding to $|\Psi\rangle_{R}$ is represented as a perturbation around the 
background given in Eqs.~\eqref{4Dbase}, \eqref{beta0}, \eqref{omega0} and \eqref{Z12nh}; the fields of the perturbation\footnote{In order to obtain~\eqref{pert2RR} one needs to
  use also $Q=R a$ following from~\eqref{Q1Q5Ra} in the case
  $Q_1=Q_5$.} read
\begin{subequations}
\begin{align}
\label{pert1RR}
w &= \frac{c_l}{Q}\,e^{-2\,i \,l\,\phi}\,\frac{\sin^{2l}\theta}{(r^2+a^2)^l}\,,\\\label{pert2RR}
B&= \frac{c_l}{2}\,e^{-2\,i
  \,l\,\phi}\,\frac{\sin^{2l}\theta}{(r^2+a^2)^l}\,\Bigl[-\frac{r^2 +
  a^2\cos^2\theta}{Q^2}\, \d t\wedge \d y -
\frac{i}{R}\,\frac{r}{r^2+a^2}\,\d r\wedge \d y \\ \nonumber
&- \cos^2\theta \left(\d\phi\wedge \d\psi - 
\frac{a}{Q} \left(\d t \wedge \d\psi + \d\phi\wedge \d y
\right)\right) - i \frac{\cos\theta}{\sin\theta} \left(\d\theta\wedge
\d\psi- \frac{a}{Q} \d\theta \wedge \d y\right) \Bigr]\,.
\end{align}
\end{subequations}

\vspace{0.2cm}

The idea of~\cite{Mathur:2003hj} is that, by a sequence of transformations in the chiral
algebra of the CFT,  $|\Psi\rangle_{NS}$ (and so
the RR state $|\Psi\rangle_{R}$ obtained after an inverse spectral
flow) can be related to a state in the RR sector of the CFT carrying
one unit of momentum. In particular, one can consider the state
$J_0^{-}|\Psi\rangle_{NS}$: it has
\be
j_{NS}=l-1\,,\quad h_{NS}=l\,,\quad \bar j_{NS}=\bar h_{NS}=l\,,
\ee
and hence is not a chiral primary. If one performs an inverse spectral
flow transformation one then reaches a RR state which is not a ground
state and whose quantum numbers are
\be
j_R=l-1\,,\quad h_R =1\,,\quad \bar j_R=l\,,\quad \bar h_R =0\,.
\ee
This state can be identified with $J_{-1}^{-}|\Psi\rangle_{R}$; it
carries momentum
\be
p = h_R-\bar h_R = 1\,.
\ee
Thanks to the fact that the operator $J_0^-$ can be identified with an
infinitesimal rotation in $\mathbb{R}^4$, it is straightforward to
generate (in the near-horizon limit) the gravity solution dual of the
state $J_{-1}^{-}|\Psi\rangle_{R}$: one starts with the solution
(\ref{pert1})-(\ref{pert2}), performs the infinitesimal rotation
associated with $J_0^-$, and finally the change of
coordinates~(\ref{inversespectralflow}). The resulting geometry, whose
explicit expression can be found in Eqs.~(3.21)-(3.30)
of~\cite{Mathur:2003hj}, solves by construction the linearized
equations in the ``near-horizon'' region. In order to
construct a real microstate of the three-charge black hole, this
``near-horizon'' geometry should be glued back to the asymptotically flat
region. This step was performed in~\cite{Mathur:2003hj} only
approximately, through a perturbative expansion in the parameter
$\epsilon= \frac{a}{\sqrt{Q}}$ (the regime $\epsilon\ll 1$ describes
geometries with a very long throat). In the next subsection we will
show that, by embedding the solution into our general
ansatz~(\ref{ansatzsummary}), the extension from the asymptotically AdS solution
to an asymptotically flat one is straightforward.  The same
formalism will also make the generalization from a linearized to an
exact background more transparent.

\subsection{Embedding MSS in our ansatz}

Let us go back to the original two-charge
geometry~\eqref{pert1RR}--\eqref{pert2RR} corresponding to the CFT
state in the RR sector $|\Psi\rangle_{R}$: to embed this 6D
supergravity solution into our 10D ansatz one first needs to specify a
10D uplift. This uplift is not unique, but we focus here on one
based on the following identifications\footnote{To match the
  conventions of \cite{Mathur:2003hj} with ours, one also needs to
  reverse the orientation: $*_6\to -*_6$, where, in our conventions,
  $\epsilon_{ty1234}=+1$.}
\be\label{pert0}
C \equiv - C_2\,,\quad w \equiv  2\,C_0 \,,\quad B\equiv B\,.
\ee
With these identifications, the state $|\Psi\rangle_{R}$ is represented by the original Lunin-Mathur geometry  given in (\ref{4Dbase}), (\ref{beta0}), (\ref{omega0}) and
(\ref{Z12nh}), plus the 10D uplift of the perturbation Eqs.~\eqref{pert1RR}--\eqref{pert2RR}:
\begin{subequations}
\begin{align}
 \label{C0R0}
C_0 &= \frac{c_l}{2 Q}\,e^{-2\,i \,l\,\phi}\,\frac{\sin^{2l}\theta}{(r^2+a^2)^l}\,,\\
B &= \frac{c_l}{2}\left\{e^{-2\,i \,l\,\phi}\,\frac{\sin^{2l}\theta}{(r^2+a^2)^l}\,\Bigl[-\frac{r^2+a^2\cos^2\theta}{Q^2} \,(\d u+\omega)\wedge (\d v+\beta)\right.\nonumber\\
\label{B2R0} &-\frac{r^2+a^2}{r^2+a^2\cos^2\theta}\,\cos^2\theta\, \d\phi\wedge
\d\psi - i \frac{\cos\theta}{\sin\theta}\,\d\theta\wedge \d\psi\Bigr] \\ \nonumber
&+\left. e^{-2\,i \,l\,\phi}\,\frac{\sin^{2l}\theta}{(r^2+a^2)^l}\,\Bigl[\frac{a}{Q}\Bigl(\d\phi + i \frac{\cos\theta}{\sin\theta} \,\d\theta -i \frac{r}{r^2+a^2}\,\d r\Bigr)\wedge \d y\Bigr]\right\}\,,
\end{align}
\end{subequations}
with $\beta$ and $\omega$ given in (\ref{beta0}) and
(\ref{omega0}). The term in the last line is $\d$-trivial and can be
gauged away. Notice that the 
perturbation~\eqref{C0R0}-\eqref{B2R0} can be embedded, at linear order in   $c_l$, in the ansatz~\eqref{ansatzsummary},  by choosing
\begin{subequations}
\begin{align}
\label{Z42charge}
Z_4&=\frac{c_l}{2}\,e^{-2\,i \,l\,\phi}\,\frac{\sin^{2l}\theta}{(r^2+a^2)^l(r^2+a^2 \cos^2\theta)}\,,\quad a_4=0\,,\\\label{delta22charge}
\delta_2&= - \frac{c_l}{2}\,e^{-2\,i \,l\,\phi}\,\frac{\sin^{2l}\theta}{(r^2+a^2)^l}\,\Bigl[\frac{r^2+a^2}{r^2+a^2\cos^2\theta}\,\cos^2\theta\,\d\phi\wedge \d\psi + i \frac{ \cos\theta}{\sin\theta}\,
\d\theta\wedge \d\psi\Bigr].
\end{align}
\end{subequations}
It is immediate to check that the $Z_4$ and $\delta_2$
above satisfy the only  supergravity constraint (\ref{eqZ4})
that is non-trivial for this solution, i.e.
\be
*_4 \d Z_4 = \d\delta_2\,.
\ee  

Since all two-charge solutions are known and are associated with a curve
$g_A(v')$ ($A=1,\ldots,8$) in $\mathbb{R}^4\times T^4$,  the solution (\ref{Z42charge})-(\ref{delta22charge}) is defined
by a particular $g_A(v')$. As mentioned before, this curve represents
the profile of the string in the duality frame in which the charges
are fundamental string and momentum: the solutions corresponding to
curves in $\mathbb{R}^4$ are the Lunin-Mathur geometries
\cite{Lunin:2001fv, Lunin:2001jy}, while general curves where
considered in \cite{Lunin:2002iz,Kanitscheider:2007wq}. Choosing a
generic profile in $\mathbb{R}^4\times T^4$ for the vibrating string breaks the rotation
invariance on $T^4$; however, when going from the F1-P to the D1-D5
duality frame, one of the directions of $T^4$, that we take to be the
direction $A=5$, is singled out. In particular a profile $g_A(v')$ that has only component 5 in $T^4$ will be a scalar in the D1-D5 frame, while
a profile in the other $T^4$ direction will correspond to a three-form. Then,  D1-D5 geometries whose profile has
components only along the directions $A=1,\dots,5$ are $T^4$ isotropic
and have, generically, all type IIB fields excited. The solution dual
to the state $|\Psi\rangle_{R}$ is exactly of this form, and indeed
one can see that it coincides with the D1-D5 geometry associated with
the curve
\be\label{profileb} 
g_1(v') = a \cos \frac{2\pi\,v'}{L} \,,\quad
g_2(v') = a \cos \frac{2\pi\,v'}{L} \,,\quad 
g_5(v') =-i\, b\,e^{-\frac{4\pi i\,l\,v'}{L}}\,, 
\ee 
at first order in $b$ and with all remaining components of $g$ set to
zero. The relation between the parameters $b$ and $c_l$ is
\be 
\frac{c_l}{2} = - b\, a R \,.  
\ee

\vspace{0.2cm}

We can embed in our ansatz also the solution in Eqs.~(3.21)-(3.30)
of~\cite{Mathur:2003hj} corresponding to the three-charge CFT state
$J_{-1}^{-}|\Psi\rangle_{R}$. As discussed in the previous section
this configuration is generated by combining the change of coordinates
corresponding to the CFT spectral flow and  $\mathbb{R}^4$
rotations. The solution carrying one unit of momentum is described by
the following geometric data\footnote{We exploit the gauge invariance
\be
a_4 \to a_4 - \dot{\lambda}^{(1)}\,,\quad \delta_2 \to \delta_2+ \mathcal{D} \lambda^{(1)}\,, 
\ee
with $\lambda^{(1)}$ a 1-form, to set $a_4=0$.}
\begin{subequations}
\begin{align}
\label{Z43charge}
Z_4&= - c_l\,l\,e^{-i\sqrt{2}\,\frac{v}{R}}\,e^{-2\,i \,l\,\phi+i\,(\phi+\psi)}\, \frac{\sin^{2l-1}\theta\,\cos\theta}{(r^2+a^2)^l\,(r^2+a^2\cos^2\theta)}\,,\\\label{delta23charge}
\delta_2&=- c_l\,l\,e^{-i\sqrt{2}\,\frac{v}{R}}\,e^{-2\,i \,l\,\phi+i\,(\phi+\psi)}\, \frac{r\,\sin^{2l-1}\theta}{(r^2+a^2)^l}\,\Bigl[\sin\theta\Bigl(\frac{\d r\wedge \d\theta}{r^2+a^2} + \frac{r\sin\theta\cos\theta}{r^2+a^2\cos^2\theta}\,\d\phi\wedge \d\psi\Bigr)\nonumber\\
& -i\,\Bigl(\frac{\cos\theta}{r^2+a^2}\,\d r\wedge \d\psi + \frac{\sin\theta}{r}\,\d\theta\wedge \d\phi\Bigr) \Bigr]\,.
\end{align}
\end{subequations} 
The other data describing the geometry 
remain, of course, unchanged at  linear order and the only non-trivial
supergravity equations are~\eqref{eqZ4a4}
\be
\dot{\delta}_2 = *_4 \dot{\delta}_2\,,\quad *_4 \mathcal{D} Z_4 = \mathcal{D}\delta_2\,,
\ee
which are easily verified. 

The advantage of rewriting the ``near-horizon'' MSS solution
 in the form of our ansatz is that the
extension to the asymptotically flat region is now immediate: it is
enough to re-add the ``1'' to the functions $Z_1$ and $Z_2$. As the
equations for $Z_4$ and $\delta_2$ do not involve $Z_1$ an $Z_2$
(when, as in our case, $\beta$ is $v$-independent), it is evident that
sending $Z_{1,2}\to Z_{1,2}+1$ does not change $Z_4$ and
$\delta_2$. Moreover, inspection of the other supergravity constraints
immediately shows that all the other geometric data are unmodified at
first order in the perturbation parameter $c_l$.
In conclusion, the geometry given by the data
in~(\ref{4Dbase})-(\ref{omega0}) and
in~(\ref{Z43charge})-(\ref{delta23charge}) solves the supergravity
equations at first order in $c_l$ and interpolates between flat space
and the AdS region for any value of
$\epsilon=\frac{a}{\sqrt{Q}}$. Thus it represents a ``perturbative''
microstate of the three-charge black hole with one unit of momentum.


\subsection{The non-linear completion}

If one wants to describe microstates that carry a macroscopic amount
of momentum charge (rather than just one quantum) one needs to take
into account higher order contributions in the perturbation
parameter $c_l$. The solution of the previous subsection fails to
solve the supergravity equations at all orders in $c_l$: terms of
order $c_l^2$ appear on the r.h.s. of the equations involving
$\mathcal{F}$ and $\omega$ (\ref{eqcalFomega}) and thus these metric coefficients
will necessarily be modified at that order.  Note that the appearance
of a non-vanishing $\mathcal{F}$ is expected for geometries carrying a
finite amount of momentum charge.

One can construct a fully non-linear three-charge solution by starting
from an exact two-charge geometry and applying a sequence of
transformations similar to the ones described above. We will give here
a sketch of the construction, specifying for simplicity to the case
$l=\frac{1}{2}$.

\vspace{0.2cm} 

Let us consider the non-linear extension of the two-charge
microstate described by the curve in (\ref{profileb}). First of all,
when working at the non-linear level one has to use real expressions
(as the trick of taking the real part of the final solution does not
apply when the equations are non-linear). There are of course many
possible curves that reduce to the real part of (\ref{profileb}) at
linear order in $b$. We will make here the following choice
\be\label{profilebbis}
g_1(v') = a \cos\left( \frac{2\pi\,v'}{L}\right) \,,\quad 
g_2(v') = a \sin\left( \frac{2\pi\,v'}{L}\right) \,,\quad 
g_5(v') = - b \sin \left(\frac{2\pi\,v'}{L}\right)\,,
\ee 
while all other components of the profile are trivial.
The exact two-charge geometry corresponding to this profile can be
derived thanks to the results of \cite{Kanitscheider:2007wq} and is
given, in our notations, by\footnote{From now on we work with
  arbitrary D1 and D5 charges; of course, in order to find the results
  of the previous two subsections one needs to use explicitly the
  constraint $Q_1=Q_5=Q$.}
\begin{subequations}
\begin{align}
\d s^2_4 &= (r^2+a^2 \cos^2\theta)\Bigl(\frac{\d r^2}{r^2+a^2}+ \d\theta^2\Bigr)+(r^2+a^2)\sin^2\theta\,\d\phi^2+r^2 \cos^2\theta\,\d\psi^2\,,\\
\beta &=  \frac{R\,a^2}{\sqrt{2}\,(r^2+a^2 \cos^2\theta)}\,(\sin^2\theta\, \d\phi - \cos^2\theta\,\d\psi)\,,\\
Z_1 &= 1+\frac{R^2}{Q_5} \frac{a^2+\frac{b^2}{2}}{r^2+a^2 \cos^2\theta}+\frac{R^2\, a^2\, b^2}{2\,Q_5}\,\frac{\cos2\phi\,\sin^2\theta}{(r^2+a^2 \cos^2\theta)(r^2+a^2)}\,,\\
Z_2 &=  1+\frac{Q_5}{r^2+a^2 \cos^2\theta}\,,\quad a_1=0\,,\\
Z_4 &=  R\, a\, b\,\frac{\cos\phi\,\sin\theta}{\sqrt{r^2+a^2}\,(r^2+a^2 \cos^2\theta)}\,,\quad a_4 =0\,,\\
\delta_2 &= \frac{ -R\, a\, b\ \sin\theta}{\sqrt{r^2+a^2}}\,\Bigl[\frac{r^2+a^2}{r^2+a^2 \cos^2\theta}\cos^2\theta \cos\phi\, \,\d\phi\wedge \d\psi + \sin\phi\, \frac{\cos\theta}{\sin\theta}\,
\d\theta\wedge \d\psi \Bigr]\,,\\
\omega &=  \frac{R\,a^2}{\sqrt{2}\,(r^2+a^2 \cos^2\theta)}\,(\sin^2\theta\,\d\phi + \cos^2\theta\,\d\psi)\,, \\
\mathcal{F} &=  0\,.
\end{align}
\end{subequations}
The relation~\eqref{Q1Q5Ra} between the $y$ radius $R$ and the charges
is now modified as
\be\label{Q1Q5Rabis}
R = \sqrt{\frac{Q_1 Q_5}{a^2+\frac{b^2}{2}}}\,.
\ee

\vspace{0.2cm}

One can now take the ``near-horizon'' limit\footnote{In the
  ``near-horizon'' limit both $a^2$ and $b^2$ are much smaller than
  $\sqrt{Q_1 Q_5}$.}, which amounts to replacing  $Z_{1,2}\to Z_{1,2}-1$, do a
spectral flow to the NSNS sector, make a finite rotation in
$\mathbb{R}^4$, and finally spectral flow back to the RR sector.  One
generates in this way a solution describing a three-charge microstate in the ``near-horizon''  region.
 The solution is
\begin{subequations}
\begin{align}
\label{3chargenhfirst}
\d s^2_4 &=  (r^2+a^2 \cos^2\theta)\Bigl(\frac{\d r^2}{r^2+a^2}+\d\theta^2\Bigr)+(r^2+a^2)\sin^2\theta\,\d\phi^2+r^2 \cos^2\theta\,\d\psi^2\,,\\
\beta &=  \frac{R\,a^2}{\sqrt{2}\,(r^2+a^2 \cos^2\theta)}\,(\sin^2\theta\, \d\phi - \cos^2\theta\,\d\psi)\,,\\
Z_1 &=  \frac{R^2}{Q_5} \frac{a^2+\frac{b^2}{2}}{r^2+a^2 \cos^2\theta}+\frac{R^2\, a^2\, b^2}{2\,Q_5}\,\cos2\hat v\,\frac{\cos^2\theta}{(r^2+a^2 \cos^2\theta)(r^2+a^2)}\,,\\
Z_2 &=  \frac{Q_5}{r^2+a^2 \cos^2\theta}\,,\quad a_1=0\,,\\
Z_4 &=  R\, a\, b\,\cos\hat v \,\frac{\cos\theta}{\sqrt{r^2+a^2}\,(r^2+a^2 \cos^2\theta)}\,,\quad a_4 =0\,,\\
\delta_2 &= R\, a\, b\,\frac{r}{\sqrt{r^2+a^2}}\,\Bigl[\cos\hat v\,\sin\theta\,\Bigl(\frac{\d r\wedge \d\theta}{r^2+a^2}+\frac{r\,\sin\theta\cos\theta}{r^2+a^2 \cos^2\theta}\,\d\phi\wedge 
\d\psi\Bigr)\nonumber\\
&\qquad - \sin\hat v \Bigl(\frac{\cos\theta}{r^2+a^2}\,\d r\wedge \d\psi + \frac{\sin\theta}{r}\,\d\theta\wedge \d\phi\Bigr)\Bigr]\,,\\
\omega &= \frac{R\,a^2}{\sqrt{2}\,(r^2+a^2 \cos^2\theta)}\,(\sin^2\theta\, \d\phi + \cos^2\theta\, \d\psi)\nonumber\\
&\qquad +\frac{R\,b^2}{\sqrt{2}}\frac{(r^2+a^2)\sin^2\theta\,\d\phi+r^2\cos^2\theta\,\d\psi}{(r^2+a^2)\,(r^2+a^2\cos^2\theta)}\,,\\\label{3chargenhlast}
\mathcal{F} &=  -\frac{b^2}{r^2+a^2}\,,
\end{align}
\end{subequations}
with
\be
\hat v = \frac{\sqrt{2}\,v}{R} -\psi\,. 
\ee
Note that the sequence of spectral flows and rotations in general mixes
the coordinates $u$ and $v$ with the $\mathbb{R}^4$ coordinates; thus,
one would have expected that the 4D metric $\d s^2_4$ and the 1-form
$\beta$ would have been modified by the series of change of
coordinates performed. However, this has not happened for the
particular transformations that generate the geometry corresponding to
the state $J_{-1}^{-}|\Psi\rangle_{R}$: at the end $\d s^2_4$ is still
flat and $\beta$ is still $v$-independent. Note also that $a_1$ is
still vanishing and $Z_2$ is still $v$-independent, which implies that
$\Theta_1=0$. One has instead a non-vanishing $\Theta_2$, that can be
computed to be
\begin{equation}
\begin{aligned}
\Theta_2&= -\frac{\sqrt{2}\,R\,a^2\,b^2}{Q_5}\,\frac{r\,\cos\theta}{r^2+a^2}\Bigl[\sin2\hat v\,\sin\theta\,\Bigl(\frac{\d r\wedge \d\theta}{r^2+a^2}+\frac{r\,\sin\theta\cos\theta}{r^2+a^2 \cos^2\theta}\, \d\phi\wedge \d\psi\Bigr) \\
& \qquad +\cos2\hat v\Bigl(\frac{\cos\theta}{r^2+a^2}\,\d r\wedge \d\psi + \frac{\sin\theta}{r}\, \d\theta\wedge \d\phi\Bigr) \Bigr]\,.
\end{aligned}
\end{equation}
Being generated from a regular solution via a globally defined
sequence of coordinate transformations, the geometry
(\ref{3chargenhfirst})-(\ref{3chargenhlast}) is guaranteed to solve
the supergravity constraints and to be regular.

The final task is the extension of the above ``near-horizon'' solution
to one that has flat as\-ymp\-toti\-cs. This task is complicated in
this case by two factors: first, the replacement $Z_{1,2}\to
Z_{1,2}+1$ generates on the r.h.s. of Eqs.~(\ref{eqomega}) and~(\ref{eqcalF}) terms
proportional to $\Theta_2$ and to the $v$-derivatives of $Z_1$ and
forces $\omega$ and/or $\mathcal{F}$ to be corrected  in order to preserve the
supergravity constraints. Second, the necessary corrections to
$\omega$ spoil the regularity of the geometry and have to be further
compensated by corrections of order $\frac{a^2}{Q_5}$. We will not
attempt to explain here a systematic technique to address these
problems. The recent results of \cite{Niehoff:2013kia} have provided a
general method to produce $v$-dependent solutions of the supergravity
equations precisely in the situation that is relevant for our problem,
i.e. when the 4D metric $\d s^2_4$ and $\beta$ are $v$-independent. The
construction of \cite{Niehoff:2013kia} applies to the restricted
ansatz in which $Z_4=a_4=\delta_2=0$, but its extension to our more
general set up is straightforward. We have checked that the solution
(\ref{3chargenhfirst})-(\ref{3chargenhlast}) fits into the scheme of
\cite{Niehoff:2013kia}, and we have used this observation to generate
the corresponding asymptotically flat geometry.  We leave the details,
as well as further applications, to a forthcoming work. We quote here,
for completeness, the final solution:
\begin{subequations}
\begin{align}
\label{3chargefinfirst}
\d s^2_4 =& \,(r^2+a^2 \cos^2\theta)\Bigl(\frac{\d r^2}{r^2+a^2}+\d\theta^2\Bigr)+(r^2+a^2)\sin^2\theta\,\d\phi^2+r^2 \cos^2\theta\,\d\psi^2\,,\\
\beta=& \, \frac{R\,a^2}{\sqrt{2}\,(r^2+a^2 \cos^2\theta)}\,(\sin^2\theta\,\d\phi - \cos^2\theta\,\d\psi)\,,\\
Z_1=&\, 1+ \frac{R^2}{Q_5} \frac{a^2+\frac{b^2}{2}}{r^2+a^2 \cos^2\theta}+\frac{R^2\, a^2\, b^2}{2\,(Q_5+a^2)}\, \frac{\cos2\hat v\,\cos^2\theta}{(r^2+a^2 \cos^2\theta)(r^2+a^2)}\,,\\
Z_2=& \,1 + \frac{Q_5}{r^2+a^2 \cos^2\theta}\,,\quad a_1=0\,,\\
Z_4=&\, R\, a\, b\,\cos\hat v \,\frac{\cos\theta}{\sqrt{r^2+a^2}\,(r^2+a^2 \cos^2\theta)}\,,\quad a_4 =0\,,\\
\delta_2=& R\, a\, b\,\frac{r}{\sqrt{r^2+a^2}}\,\Bigl[\cos\hat v\,\sin\theta\,\Bigl(\frac{\d r\wedge \d\theta}{r^2+a^2}+\frac{r\,\sin\theta\cos\theta}{r^2+a^2 \cos^2\theta}\,\d\phi\wedge 
\d\psi\Bigr)\nonumber\\
&\qquad - \sin\hat v \Bigl(\frac{\cos\theta}{r^2+a^2}\,\d r\wedge \d\psi + \frac{\sin\theta}{r}\,\d\theta\wedge \d\phi\Bigr)\Bigr]\,,
\end{align}
\begin{align}
\omega &=  \frac{R\,a^2}{\sqrt{2}\,(r^2+a^2 \cos^2\theta)}\,(\sin^2\theta\,\d\phi + \cos^2\theta\,\d\psi)\nonumber\\
&+\frac{R\,b^2}{\sqrt{2}}\frac{(r^2+a^2)\sin^2\theta\,\d\phi+r^2\cos^2\theta\,\d\psi}{(r^2+a^2)\,(r^2+a^2\cos^2\theta)}\nonumber\\
&-\frac{R\,a^2\,b^2}{2\sqrt{2}\,(Q_5+a^2)}\,\Bigl[\cos2\hat v\,\frac{a^2 \sin^2\theta\, \d\phi-r^2 \,\d\psi}{(r^2+a^2)\,(r^2+a^2\cos^2\theta)}\,\cos^2\theta\nonumber\\
&+\sin2\hat v\,\frac{r \cos\theta\,\d r - (r^2+a^2)\sin\theta\, \d\theta}{(r^2+a^2)^2}\,\cos\theta\Bigr] \,,\\\label{3chargefinlast}
\mathcal{F} &= -\frac{b^2}{r^2+a^2}\,.
\end{align}
\end{subequations}
This geometry solves the non-linear supergravity equations, is regular
and horizon-less, and at large distances reduces to the geometry of
the black hole with D1, D5 and P charges.

\vspace{1cm}

\centerline{\large \bf Acknowledgements}

\vspace{0.5cm}

\noindent  
We thank I. Bena, G. Dall'Agata, G. Gibbons, S. Mathur, M. Shigemori, A. Tomasiello, D. Turton and N. Warner for several enlightening discussions. We especially thank I. Bena for useful comments on the draft. SG and LM have been partially supported by MIUR-PRIN contract 2009-KHZKRX, by the Padova University Project CPDA119349 and by INFN. RR has been partially supported by STFC Standard Grant ST/J000469/1 `String Theory, Gauge Theory and Duality'. SG, MP and RR have 
been partially supported by the CNRS grant PICS "Aspects of String Theory with Fluxes". MP wishes to thank the Theory Group of Imperial College, London, for hospitality
during the completion of this work.   RR wishes to thank Universit\'e Pierre et Marie Curie (Paris), Universit\`a di Torino and Coll\`ege de France for hospitality, and the `Institut Lagrange de Paris' for support during the completion of this work. 

\vspace{1cm}

\begin{appendix}

\section{Conventions}
\label{app:conv}

In this paper we will consider ten dimensional Minkowski space we use the signature $(-,+,\ldots,+)$, as well as eight- and four-dimensional euclidian 
space.  We distinguish local flat indices by underlying them. 

In a $D$-dimensional space, take an oriented vielbein $e^{\ul m}$, with ${\ul m}=0,\ldots, D-1$  if the space is Minkowskian and ${\ul m}=1,\ldots, D$ if it is  euclidean. Then the Hodge-$*$ of a $k$-form $A_k$ is defined as follows:
\be
*A_k\equiv \frac{1}{k!(D-k)!}\, \epsilon_{\ul{m}_1\ldots \ul{m}_{D-k}\ul{m}_{D-k+1}\ldots\ul{m}_{D}}A^{\ul{m}_{D-k+1}\ldots\ul{m}_{D}}\, e^{\ul{m}_1}\wedge \ldots \wedge  e^{\ul{m}_{D-k}}\,,
\ee
where $\epsilon_{\ul{m}_1\ldots \ul{m}_{D}}$ is the totally antisymmetric symbol, such that $\epsilon_{\ul{0}\ldots \ul{D-1}}=1$ and $\epsilon_{\ul{1}\ldots \ul{D}}=1$ in Minkowskian and Euclidean spaces respectively.

\bigskip

We often use the operator $\lambda$ which acts on a $k$-form $A_k$ as follows
\be
\label{lambdat}
\lambda(A_k)=(-)^{k(k-1)/2}A_k\,,
\ee
that is, $\lambda$ exchanges the order of the indices of the form it is action on.

\bigskip

We define the contraction of a $k$-form $A_k$ by a vector $X=X^m\partial_m$  by 
\be
\iota_XA_k
\equiv \frac{1}{(k-1)!}X^mA_{mn_1\ldots n_{k-1}}\d x^{n_1}\wedge\ldots\d x^{n_{k-1}}\,,
\ee 
and we use  the shorthand notation $\iota_m\equiv\iota_{\partial_m}$.

We also make use of the full contraction of two $k$-forms,  $A_k$ and $B_k$, 
\be
A_k\cdot B_k\equiv \frac{1}{p!} A_{M_1\ldots M_p}B^{M_1\ldots M_p}\,,
\ee
which is generalised to the contraction of polyforms $A=\sum_k A_k$ and $B=\sum_k B_k$ as
\be
A\cdot B\equiv \sum_k A_k\cdot B_k \, . 
\eeq

When acting on spinors, a $k$-form is implicitly taken to be contracted by gamma matrices
\be
\label{gammacont}
A_{k} \equiv \frac{1}{k!}A_{m_{1} \ldots m_k}\gamma^{m_{1} \ldots m_k}\,,
\ee
where, as usual, $\gamma^{m_1\ldots m_k}\equiv \gamma^{[m_1}\ldots\gamma^{m_k]}$.

\bigskip

We use spinor in 10 Minkowskian and 8 and 4 Euclidean dimensions. The corresponding chirality operators are
\be
\Gamma_{(10)}=\Gamma^{\ul{0\ldots 9}}\,,\qquad \gamma_{(8)}=\gamma^{\ul{1\ldots 8}}\,,\qquad\gamma_{(4)}=\gamma^{\ul{1\ldots 4}}\,.
\ee
When we consider the split from 10 to 2+8 dimensions we use the following gamma matrices decomposition
\be\label{gamma2+8bis}
\Gamma^{\ul 0}=\ii \sigma_2\otimes \gamma_{(8)}\, ,\qquad \Gamma^{\ul 1}=\sigma_1\otimes \gamma_{(8)}\, ,\qquad  \Gamma^{\ul a}=\bbone\otimes \gamma^{\ul a}\,.
\ee
We can also split the ten-dimensional charge conjugation matrix 
\be
C_{(10)}=\ii\sigma_2\otimes C_{(8)}\gamma_{(8)}\,,
\ee
where  $C_{(8)}$ is such that $C_{(8)}\gamma_{\ul a}C_{(8)}^{-1}=\gamma^T_{\ul a}$.  
The Majorana condition on a ten-dimensional spinor $\epsilon$ imposes that 
\be
\bar\epsilon\equiv \epsilon^\dagger\Gamma^{\ul{0}}=\epsilon^TC_{(10)} \, , 
\ee
 and on an eight-dimensional spinor $\eta$ imposes 
 \be
 \eta^\dagger=\eta^TC_{(8)} \, .
 \ee
By choosing $C_{(8)}=\bbone$ and then $C_{(10)}=\Gamma^{\ul0}$ one obtains the real representation.

We also consider the split of the Eucliden 8-dimensions into 4+4 dimensions as in (\ref{gamma4+4}), under which
\be
C_{(8)}=C_{(4)}\otimes \hat C_{(4)}\,,
\ee 
where $C_{(4)}$ (and analogously $\hat C_{(4)}$) now satisfies the identities $C_{(4)}\gamma_{i}C^{-1}_{(4)}=-\gamma^T_{i}$. 
For instance, we could take the following explicit representation for four-dimensional gamma matrices
\be\label{repr1}
\gamma^{\ul 1}=\sigma_1\otimes \bbone\,,\quad~~~
\gamma^{\ul 2}=\sigma_2\otimes \bbone\,,\quad~~~
\gamma^{\ul 3}=\sigma_3\otimes \sigma_2\,,\quad~~~
\gamma^{\ul 4}=\sigma_3\otimes \sigma_1
\ee
and then $\gamma_{(4)}=\sigma_3\otimes\sigma_3$. In this basis we can choose $C_{(4)}=\sigma_2\otimes\sigma_1$.

\bigskip

In this paper we use the democratic formulation \cite{Bergshoeff:2001pv} of type II supergravities in the conventions spelled out in detail in Appendix A of \cite{Lust:2008zd}, up to a sign flip $H\rightarrow -H$.
Let us just recall some informations relevant for the analysis of supersymmetry. 

For the Ramond-Ramond fields we consider the full sum of field strenghts
\beq
\label{RRdefbis}
F = \sum_k F_{k} \, , 
\eeq
with  $k$ even  (from 0 to 10) in IIA and odd (from 1 to 9) in IIB. The redundant degrees of freedom  in $F$ are eliminated by the self-duality constraint
\beq
\label{10dselfd}
F =  \ast \lambda(F)  \, . 
\eeq
where $\lambda$  is given in \eqref{lambdat}.  

\vspace{0.2cm}

The fermionic content of type II supergravity consists of a doublet of gravitino's and dilatino's
\be
\psi_M=(\psi^1_{M},\psi^2_M) \qquad \,,\qquad \lambda=(\lambda^1,\lambda^2) \, .
\ee
The components of the doublet have  different chirality in type IIA and  the same chirality in type IIB.  In both theories we fix the chirality to be
\be
\Gamma_{(10)}\psi^1_{M}=\psi^1_{M} \quad\,,\quad  \Gamma_{(10)}\lambda^1=-\lambda^1\, .
\ee
The type II supersymmetry transformations are parameterized by a doublet of Majorana-Weyl spinors $\epsilon=(\epsilon_1,\epsilon_2)$,  of opposite chirality in IIA and same
chirality in IIB
\be
\Gamma_{(10)}\epsilon_1=\epsilon_1\quad \mbox{(IIA)}\,,  \qquad \Gamma_{(10)}\epsilon_2=\mp\epsilon_2  \quad \mbox{(IIB)}\,. 
\ee
In our conventions, the type II supersymmetry transformations of \cite{Bergshoeff:2001pv} can be written as follows
\begin{subequations}\label{susyvargrav}
\begin{align}
\label{susyvargrav1}
\delta\psi^{1}_M & =\, \Big(\nabla_M - \frac14\iota_M H \Big)\epsilon_1 +\frac{1}{16}e^\phi
F\,\Gamma_M\Gamma_{(10)}\epsilon_2\ , \\
\label{susyvargrav2}
 \delta\psi^{2}_M &=\,  \Big(\nabla_M + \frac14\iota_MH \Big)\epsilon_2-\frac{1}{16}e^\phi
\lambda(F)\,\Gamma_M\Gamma_{(10)}\epsilon_1\ , \\
\label{susyvardil1}
 \delta\lambda^{1} &= \,  \left(\d\phi  - \frac12 H\right)\epsilon_1+\frac{1}{16}e^{\phi}\Gamma^M F\,\Gamma_M\Gamma_{(10)}\epsilon_2\ , \\
\label{susyvardil2}
 \delta\lambda^{2} &=\,  \Big(\d\phi + \frac12 H\Big)\epsilon_2-\frac{1}{16}e^{\phi}\Gamma^M \lambda(F)\,\Gamma_M\Gamma_{(10)}\epsilon_1\ ,
\end{align}
\end{subequations}
where fluxes are contracted on gamma matrices as in \eqref{gammacont}.

\section{Spin(7) structures}
\label{app:spin7}

Take an 8-dimensional space $X$ and a globally defined, nowhere vanishing Majorana-Weyl  spinor $\eta$. The structure group oh the spin bundle  is reduced from Spin(8) to Spin(7).
We choose the charge conjugation matrix $C_{(8)}=\bbone$, so that all gamma matrices $\gamma_a$ are real,  $\eta$ is real and satisfies $\gamma_{(8)}\eta=\eta$, with $\gamma_{(8)}=\gamma_{\ul{1\ldots 8}}$. 
Furthermore we normalize $\eta$ in such a way that
\be\label{appnorm}
\eta^T\eta=1 \,  . 
\ee
An equivalent way of defining a reduced Spin(7) structure is via a four-form 
\be
\Omega=\frac1{4!}\,\eta^T\gamma_{abcd}\eta\, \d y^a\wedge \d y^b\wedge \d y^c\wedge \d y^d \, . 
\ee
If $\eta$ has positve chirality,  then $\Omega$ is self-dual 
\be
*\Omega=\Omega  \, , 
\ee
while if   $\gamma_{(8)}\eta=-\eta$ then  $*\Omega=-\Omega$.

The different tensors can be decomposed in different representations of Spin(7). 
A vector transforms in the representation ${\bf 8}$   of Spin(7).
On the other hand, a tensor with two antisimmetric indices decomposes as follows in irreducible representations 
\be
\begin{aligned}
&\alpha_{[ab]} ={\bf 7}\oplus {\bf 21} \, .
\end{aligned}
\ee
We can use $\Omega$ to construct the corresponding projectors. In particular 
\be
\begin{aligned}
(P_{\bf 7})_{ab}{}^{cd}&=\frac14\Big(\delta^{[c}_{[a}\delta^{d]}_{b]}-\frac12\Omega_{ab}{}^{cd}\Big)\\
\end{aligned}
\ee
is the projector on  the ${\bf 7}$  of Spin(7). 

It is also useful to observe that, if $\chi$ is any four-form, then one can construct a two-form transforming in the ${\bf 7}$ of Spin(7)
\be\label{7four}
T^{\bf 7}\equiv \iota_{[a}\Omega\cdot \iota_{b]}\chi\,\d y^a\wedge \d y^b\equiv\frac{1}{ 3!}(\Omega_{[a}{}^{cde} \chi_{b]cde})\wedge\d y^a\wedge \d y^b \, .
\ee

\subsection{Decomposition of spinors}
We can decompose an arbitrary spinor $\xi$ on $X$  by using as basis $\eta, \gamma_a\eta,\gamma_{ab}\eta$. 
In particular, if $\eta$ has positive chirality, we can decompose  a positive/negative  chirality spinor $\xi_{+/-}$ as
\be\label{spindec}
\begin{aligned}
\zeta_+&=\chi^{\bf 1}\eta+\frac1{2!}\,\chi^{\bf 7}_{ab}\gamma^{ab}\eta \, , \\
\zeta_-&=\chi^{\bf 8}_a\gamma^a\eta \, ,
\end{aligned}
\ee 
where $\chi^{\bf 1}$ is a singlet of Spin(7), $\chi^{\bf 8}$ is a vector and  the two-form $\chi^{\bf 7}_{ab}$ transforms as  ${\bf 7}$ of Spin(7). This is due to the following projector conditions:
\be\label{Pgamma}
(P_{\bf 7})_{ab}{}^{cd} \gamma_{cd}\eta=\gamma_{cd}\eta\,.
\ee 
Hence $\zeta_+\in {\bf 1}\oplus {\bf 7}$ and $\zeta_-\in {\bf 8}$. Notice also that  ($\eta$, $\gamma_{ab} \eta$) and  ($\gamma^a\eta$) form a basis for the positive and
negative chirality spinors, respectively.  If $\gamma_{(8)}\eta=-\eta$, we have a similar decomposition with chiral and antichiral spinors exchanged.

The spinorial basis ($\eta$, $\gamma_a \eta$, $\gamma_{ab} \eta$) obeys the following orthogonality conditions
\be
\begin{aligned}
\eta^T\gamma_a\gamma^b\eta&=\delta^b_a \, ,  \\
\eta^T\gamma_{ab}\gamma^{cd}\eta&=-8(P_{\bf 7})_{ab}{}^{cd} \, ,
\end{aligned}
\ee
which can be used to invert the decomposition (\ref{spindec})
\be
\chi^{\bf 1}=\eta^T\zeta_+\,,\quad~~~~
\chi^{\bf 7}_{ab}=-\frac14\eta^T\gamma_{ab}\zeta_+\,,\quad~~~~
\chi^{\bf 8}_{a}=\eta^T\gamma_{a}\zeta_- \, .
\ee

Another useful identity is the following. Let us state it for a non-normalized spinor $\eta$, with $\eta^T\eta\neq 1$. If we define a (non-normalized) $\Omega_{abcd}=\eta^T\gamma_{abcd}\eta$, then 
\be\label{usid}
\eta^T\gamma_{ab}\dot\eta-\frac12\eta^T\gamma_{ab}m_{\rm A}\,\eta=-\frac1{8\eta^T\eta}\,\iota_{[a}\Omega\cdot\frac{\d}{\d v}\big(\iota_{b]}\Omega\big)\,.
\ee
Notice that all terms in this equations transform as two-forms in ${\bf 7}$, cf.~(\ref{7four}) and (\ref{Pgamma}).

\section{Integrability for null  Killing vectors}
\label{app:intnull}

In this section we discuss again the relation between supersymmetry and equations of motion for the case of null Killing vector, which is relevant for this  paper.
In the derivation we do not explicitly include branes as localised sources  and we only consider the closed string sector.

Let us first define
\be
\begin{aligned}
\cale_{MN}&=R_{MN}+2\nabla_{M}\nabla_N\phi-\frac12H_M\cdot
H_N-\frac14e^{2\phi}F_M\cdot F_N\,,\\
\calh_{MN} &=e^{2\phi}*_{10}\Big[
\d(e^{-2\phi}*_{10} H) +\frac12(*_{10} F\wedge F)_{8}
\Big]\,,\\
\mathcal{O}&=2R-H^2+8\left[\nabla^2\phi-(\d\phi)^2\right]\,,\label{dlth}
\end{aligned}
\ee
where $F$ is the sum of all RR filed-strengths as in \eqref{RRdefbis}. Using this notation, the  string-frame trace-reversed Einstein equations,  the $B$-field equations and the dilation's equation are simply
\be
 \cale_{MN}=0 \,, \qquad   \calh_{MN}=0 \,, \qquad  \mathcal{O}=0 \, .
\ee
 On the other hand,
 \be
 \label{fluxesBI}
 \d H=0  \qquad \,, \qquad  \d_HF=0  \, 
 \ee
 give  the Bianchi identity for the B-field and the  Bianchi identities and equations of motion for the RR-fields. 

\vspace{0.2cm}

The starting point of our discussion is provided by  the equations (11.4) in \cite{Lust:2008zd}.  
They immediately imply that, if the type II  background is supersymmetric,  the following equations are satisfied
\be\label{inteq}
\begin{aligned}
&\cale_{MN}\Gamma^N\epsilon_1+\frac12[\calh_{MN}\Gamma^N+\iota_M(\d H)]\epsilon_1+\frac14(\d_H F)\Gamma_{M}\Gamma_{(10)}\epsilon_2=0 \, , \\
&\cale_{MN}\Gamma^N\epsilon_2-\frac12[\calh_{MN}\Gamma^N+\iota_M(\d H)]\epsilon_2-\frac14\Gamma_{(10)}\lambda(\d_H F)\Gamma_{M}\epsilon_1=0 \, ,\\
&\mathcal{O}\epsilon_1+2(\d H) \epsilon_1- (\d_H F)\epsilon_2=0 \, , \\
&\mathcal{O}\epsilon_2-2 (\d H) \epsilon_2-\lambda(\d_{H}F) \epsilon_1=0 \, .
\end{aligned}
\ee
If we then impose the  Bianchi identities \eqref{fluxesBI}, it is easy to see from the  last two equations in \eqref{inteq} that
the dilaton equation,  $\mathcal{O}=0$,  is automatically fulfilled.  On the other hand, the first two equations of (\ref{inteq}) reduce to
 \be\label{inteq2}
\begin{aligned}
&(\cale_{MN}+\frac12\calh_{MN})\Gamma^N\epsilon_1=0 \, , \\
&(\cale_{MN}-\frac12\calh_{MN})\Gamma^N\epsilon_2=0 \, . 
\end{aligned}
\ee

These can be reduced to a set of bosonic conditions using the fact that  $\Gamma^{\ul u}\epsilon_1$  and $\Gamma^{\ul a}\epsilon_1$, with  $a=1,\ldots,8$,  form a set of linearly independent spinors\footnote{Remember that for $K$ null,  both $\epsilon_1$ and $\epsilon_2$ are annihilated by $\Gamma^{\ul v}$. } (the same is true for $\epsilon_2$)
 \be
 \label{inteq3}
\begin{aligned}
&\cale_{\ul{Mu}}+\frac12\calh_{\ul{Mu}}=0\,,\quad~~~~~~~~\cale_{\ul{Ma}}+\frac12\calh_{\ul{Ma}}=0  \, , \\
&\cale_{\ul{Mu}}-\frac12\calh_{\ul{Mu}}=0\,,\quad~~~~~~~~\cale_{\ul{Ma}}-\frac12\calh_{\ul{Ma}}=0 \, , 
\end{aligned}
\ee
where again $\ul{a}=1\,\ldots, 8$. Clearly, these are in turn equivalent to the set of equations
 \be\label{inteq4}
\begin{aligned}
&\cale_{\ul{Mu}}=0\,,\quad~~~~~~~~\cale_{\ul{Ma}}=0  \, , \\
&\calh_{\ul{Mu}}=0\,,\quad~~~~~~~~\calh_{\ul{Ma}}=0 \, .
\end{aligned}
\ee 
Since  $\calh_{MN}$ is antisymmetric, the second line of (\ref{inteq4}) is actually equivalent to the complete set of B-field equations of motion $\calh_{MN}=0$.  
The first line provides almost all components of the Einstein equations $\cale_{MN}=0$, with the exclusion of the $\ul{vv}$ component 
\be\label{vvcomp}
\cale_{\ul{vv}}=0 \, .
\ee

We then reach the conclusion that, if $K$ is {\em null}, imposing supersymmetry and the BI's (\ref{fluxesBI}) automatically implies all  remaining equations of motion but  (\ref{vvcomp}), which then needs to be checked separately.

\section{The missing supersymmetry equations}
\label{app:extrasusy}

Equations (\ref{killingmetric}) and (\ref{gensusy}) are not sufficient for guaranteeing the supersymmetry of a ten dimensional background
and  must be supplemented by additional conditions \cite{Tomasiello:2011eb}.  The problem is that some information contained in the Killing spinor equations is
projected out when going to the polyform equations.  In fact, by looking at Appendix B of  \cite{Tomasiello:2011eb} one can easily identify which are the Killing spinor equations which are not captured by (\ref{killingmetric}) and (\ref{gensusy}). The missing conditions are provided by equations (B.39) and (B.40) therein and  correspond exactly to the
component of two gravitino equations  along the null-directions $e_{+_1}$ and $e_{+_2}$.  

In our case, where $K$ is null,  we can identify $e_{+_1}$ and $e_{+_2}$ with $E_{\ul v}$ and then the missing supersymmetry conditions are given by the ${\ul v}$-component of the gravitino variations given in (\ref{susyvargrav}):
\be
\label{veq}
\begin{aligned}
&(\nabla_{\ul v}-\frac14 \iota_{\ul{v}}H)\epsilon_1+\frac{1}{16}\, e^\phi\, F\Gamma_{\ul v}\Gamma_{(10)}\epsilon_2=0 \, , \\
&(\nabla_{\ul v}+\frac14 \iota_{\ul{v}}H)\epsilon_2-\frac{1}{16}\, e^\phi\, \lambda(F)\Gamma_{\ul v}\epsilon_1=0 \, . 
\end{aligned}
\ee
Our aim is to express these equations in a more tractable form, different (but equivalent) to equations (3.1d)-(3.1e) of \cite{Tomasiello:2011eb}.

We first  split the NSNS and RR fluxes as in \eqref{fluxdec}
\be
\begin{aligned}
\label{fluxapp}
H&=h+E^{\ul u}\wedge h_{\ul{u}}+E^{\ul v}\wedge h_{\ul{v}}+E^{\ul u}\wedge E^{\ul v}\wedge h_{\ul{uv}} \, , \\
F&=f+E^{\ul u}\wedge f_{\ul{u}}+E^{\ul v}\wedge f_{\ul{v}}+E^{\ul u}\wedge E^{\ul v}\wedge f_{\ul{uv}} \, . 
\end{aligned}
\ee
Then we compute the covariant derivatives using the ansatz for the metric \eqref{metric} and the spinors \eqref{ansatz1}
\be
\begin{aligned}
\nabla_{\ul v}\epsilon_1\equiv E^M_{\ul v}\nabla_M\epsilon_1=&\left(\begin{array}{c}1 \\ 0
\end{array}\right)\otimes \Big[  e^{-2D}\dot\eta_1+\frac{1}{4}(\cald\omega+W\cald\beta)\eta_1-\frac12 e^{-2D}m_{\rm A}\eta_1  \Big]\\
&-\left(\begin{array}{c}0 \\ 1
\end{array}\right)\otimes \Big[  \frac{\sqrt{2}}{4}\Big(e^{-2D}\cald e^{2D}-\dot\beta\Big)\eta_1  
\Big] \, , 
\end{aligned}
\ee
where  we have introduced the two-form 
\be
m_{\rm A}=\frac12\,\delta_{\ul{ac}}\,\dot e^{\ul c}_d\,e^{d}_{\ul b}\, e^{\ul a}\wedge e^{\ul b}\,.
\ee 
 Analogously
\be
\begin{aligned}
\nabla_{\ul v}\epsilon_2=E^M_{\ul v}\nabla_M\epsilon_1=&
\left(\begin{array}{c}1 \\ 0
\end{array}\right)\otimes \Big[  e^{-2D}\dot\eta_2+\frac{1}{4}(\cald\omega+W\cald\beta)\eta_2-\frac12 e^{-2D}m_{\rm A}\eta_2  \Big]\\
&+\left(\begin{array}{c}0 \\ 1
\end{array}\right)\otimes \Big[  (-)^{|F|} \frac{\sqrt{2}}{4}\Big(e^{-2D}\cald e^{2D}-\dot\beta\Big)\eta_2 
\Big] \, . 
\end{aligned}
\ee

The terms in \eqref{veq} containing the fluxes are easily simplified using \eqref{fluxapp} and the chirality properties of the spinors.
Then, we can rewrite (\ref{veq}) as 
\begin{subequations}
\begin{align}
\label{remsusy1}
& (e^{-2D}\cald e^{2D}-\dot\beta)\eta_1-h_{\ul{uv}}\eta_1-\frac12e^\phi\, f\,\eta_2=0 \, , \\
\label{remsusy2}
& (e^{-2D}\cald e^{2D}-\dot\beta)\eta_2+h_{\ul{uv}} \eta_2-\frac12(-)^{|f|}e^\phi\, \lambda(f)\,\eta_1=0 \, , \\
\label{remsusy3}
& \dot\eta_1-\frac12m_{\rm A}\,\eta_1+\frac{e^{2D}}{ 4} \,(\cald\omega+W\cald\beta)\,\eta_1-\frac{e^{2D}}{4} h_{\ul v}\, \eta_1+\frac1{8}\, e^{2D+\phi}f_{\ul v}\,\eta_2=0
\, , \\
\label{remsusy4}
&  \dot\eta_2-\frac12m_{\rm A}\,\eta_2+ \frac{e^{2D}}{ 4}  (\cald\omega+W\cald\beta)\,\eta_2+\frac{e^{2D}}{4} h_{\ul v}\,\eta_2-\frac{e^{2D+\phi}}{8} \lambda(f_{\ul v})\,\eta_1=0 \, .
\end{align}
\end{subequations}

\bigskip

Each of the spinors $\eta_1$ and $\eta_2$ defines a Spin(7) structure.  We can use them to expand \eqref{remsusy1}-\eqref{remsusy4}  in a natural spinorial basis.

\vspace{0.2cm}

Let us start with  \eqref{remsusy1}. By construction all terms have opposite chirality with respect to $\eta_1$. As discussed in Appendix  \ref{app:spin7}, 
we can expand it in the basis $\gamma_a\eta_1$ with  $a=1,\ldots,8$. The different components are 
obtained by contraction it with $\eta_1^T\gamma_a$.   Using the definition of $\Phi$  (\ref{Phi}),  we can write
\be
\label{eqa1}
\sqrt{2}\sin^2\theta(e^{-2D}\cald e^{2D}-\dot\beta-h_{\ul{uv}})-\frac12e^\phi(\iota_a f\cdot \Phi+f\cdot\iota_a\Phi)\d x^a=0 \, . 
\ee
We can repeat the same procedure, expanding \eqref{remsusy2} in the basis $\gamma_a\eta_2$, 
\be
\label{eqa2}
\sqrt{2}\cos^2\theta(e^{-2D}\cald e^{2D}-\dot\beta+h_{\ul{uv}})+\frac12e^\phi(\iota_a f\cdot \Phi-f\cdot\iota_a\Phi)\d x^a=0 \, .
\ee
\eqref{eqa1} and \eqref{eqa2} combine into the following algebraic equations
\be\label{esapp}
\begin{aligned}
e^{-2D}\cald e^{2D}-\dot\beta+\cos2\theta\, h_{\ul{uv}}&=\frac1{\sqrt{2}}e^\phi\, (f\cdot\iota_a\Phi)\,\d x^a \, , \\
h_{\ul{uv}}+\cos2\theta\,(e^{-2D}\cald e^{2D}-\dot\beta)&=-\frac1{\sqrt{2}}e^\phi\, (\iota_a f\cdot\Phi)\,\d x^a \, ,
\end{aligned}
\ee
which can be further simplified if one uses the supersymmetry equations \eqref{gensusy}. Indeed \eqref{gensusya}  allows to determine
some components of the NSNS flux
\be
\label{Hsplit}
\begin{aligned}
h_{\ul u}&=\cos2\theta\, e^{2D}\cald\beta \, , \\
h_{\ul{uv}}&= \cos2\theta\, \dot\beta -e^{-2D}\cald(\cos2\theta\, e^{2D})\, ,
\end{aligned}
\ee
while \eqref{gensusyb} gives 
\be
\label{155}
\begin{aligned}
& f_{\ul u}=\sqrt{2}e^{2D-\phi}\cald\beta\wedge\Phi \, , \\
& f_{\ul{uv}}+\cos2\theta\, f=-\sqrt{2}e^{-\phi}\big[ e^{-2D+\phi}\cald(e^{2D-\phi}\Phi)-h\wedge \Phi-\dot\beta\wedge\Phi \big] \, .
\end{aligned}
\ee 

In particular, using  \eqref{Hsplit}, we can rewrite \eqref{esapp} as 
\begin{subequations}
\begin{align}
\label{esappbis}
& e^{-2D}\cald e^{2D}-\dot\beta+\cald\log(\sin2\theta)= \frac1{\sqrt{2}\,\sin^2(2\theta)}e^\phi\, (f\cdot\iota_a\Phi)\,\d x^a \, , \\
&  \cald_a \cos2\theta= \frac1{\sqrt{2}}e^\phi\, \iota_a f\cdot\Phi \,,  \\
& \frac{\d}{\d v}(\cos 2\theta)= \frac{\sqrt{2}}4e^{2D+\phi} f_{\ul v}\cdot\Phi  \, .
\end{align}
\end{subequations}

\vspace{0.2cm}

Let us now pass to the last two equations,  \eqref{remsusy3} and \eqref{remsusy4}.  We can expand them in the basis $(\eta_1,\gamma_{ab}\eta_1)$ and $(\eta_2,\gamma_{ab}\eta_2)$, respectively. Let us first contract \eqref{remsusy3} with $\eta_1^T$. Recalling that we set $\eta^T_1\eta_1=\sqrt{2}\sin^2\theta$, we get
\be
\frac{1}{\sqrt{2}}\frac{\d}{\d v}(\sin^2\theta)+\frac18e^{2D+\phi} f_{\ul v}\cdot\Phi=0 \, . 
\ee
On the other hand, contracting \eqref{remsusy4} with $\eta^T_2$  and using $\eta^T_1\eta_1=\sqrt{2}\cos^2\theta$ gives
\be
\frac{1}{\sqrt{2}}\frac{\d}{\d v}(\cos^2\theta)-\frac18e^{2D+\phi} f_{\ul v}\cdot\Phi=0 \, .
\ee
These two equations are clearly equivalent and can be rewritten as 
\be
\frac{1}{\sqrt{2}}\frac{\d}{\d v}(\cos 2\theta)-\frac14e^{2D+\phi} f_{\ul v}\cdot\Phi=0 \, .
\ee

We now have to consider the $\gamma_{ab}\eta_1$ and $\gamma_{ab}\eta_2$ components of  \eqref{remsusy3} and \eqref{remsusy4}. 
Let us introduce the four-forms $\Omega^{(1)}$ and $\Omega^{(2)}$ defined as
\be
\Omega^{(1)}_{abcd}=\eta_1^T\gamma_{abcd}\eta_1\quad\,,\quad \Omega^{(2)}_{abcd}=\eta_2^T\gamma_{abcd}\eta_2 \, .
\ee
Then, by using (\ref{usid}) we can write
\be
\begin{aligned}
\eta_1^T\gamma_{ab}\dot\eta_1-\frac12\eta_1^T\gamma_{ab}m_{\rm A}\,\eta_1&=-\frac{1}{8\sqrt{2}\sin^2\theta}\,\iota_{[a}\Omega^{(1)}\,\cdot \,\frac{\d}{\d v}\big(\iota_{b]}\Omega^{(1)}\big) \, , \\
\eta_2^T\gamma_{ab}\dot\eta_2-\frac12\eta_2^T\gamma_{ab}m_{\rm A}\,\eta_2&=-\frac{1}{8\sqrt{2}\cos^2\theta}\,\iota_{[a}\Omega^{(2)}\,\cdot \,\frac{\d}{\d v}\big(\iota_{b]}\Omega^{(2)}\big) \, .
\end{aligned}
\ee
On the other hand, by using the  projectors on the  representations ${\bf 7}$ of the two reduced Spin(7) structure group defined by $\eta_1$ and $\eta_2$, 
\be
\begin{aligned}
(P^{(1)}_{\bf 7})_{ab}{}^{cd}&=\frac14\Big(\delta^{[c}_{[a}\delta^{d]}_{b]}-\frac1{2\sqrt{2}\sin^2\theta}\Omega^{(1)}_{ab}{}^{cd}\Big) \,  , \\
(P^{(2)}_{\bf 7})_{ab}{}^{cd}&=\frac14\Big(\delta^{[c}_{[a}\delta^{d]}_{b]}-\frac1{2\sqrt{2}\cos^2\theta}\Omega^{(2)}_{ab}{}^{cd}\Big) \, , \\
\end{aligned}
\ee
we can write
\be
\begin{aligned}
& \eta_1\gamma_{ab}(\cald\omega+W\cald\beta)\eta_1- \eta_1h_{\ul v}\,\eta_1=- 4 \sqrt{2}\sin^2\theta\, (P^{(1)}_{\bf 7})_{ab}{}^{cd}(\cald\omega+W\cald\beta-h_{\ul v})_{cd } \, ,\\
& \eta_2\gamma_{ab}(\cald\omega+W\cald\beta)\eta_2+ \eta_2h_{\ul v}\,\eta_2 =- 4 \sqrt{2}\cos^2\theta\, (P^{(2)}_{\bf 7})_{ab}{}^{cd}(\cald\omega+W\cald\beta+h_{\ul v})_{cd} \, . \\
\end{aligned}
\ee
Furthermore we have
\be
\begin{aligned}
& \eta_1^T\gamma_{ab}f_{\ul v}\,\eta_2= \Phi\cdot (\gamma_{ab} f_{\ul v}) \, , \\
&  \eta_2^T\gamma_{ab}\lambda(f_{\ul v})\,\eta_1= - \Phi\cdot (  f_{\ul v}\gamma_{ab}) \, , 
\end{aligned}
\ee
where the action of gamma matrices on differential forms is explained after \eqref{alteqs} and in the present case is explicitly given by
\be
\begin{aligned}
\gamma_{ab} f_{\ul v}&=\iota_a\iota_b f_{\ul v}+\d x_a\wedge \d x_b \wedge f_{\ul v}+2\d x_{[a}\wedge \iota_{b]}f_{\ul v} \, , \\
 f_{\ul v}\gamma_{ab}&=\iota_a\iota_b f_{\ul v}+\d x_a\wedge \d x_b \wedge f_{\ul v}-2\d x_{[a}\wedge \iota_{b]}f_{\ul v} \, ,
\end{aligned}
\ee
with  $\d x_a\equiv g_{ab}\d x^b$.  Putting all these steps together, we obtain the following equations 
\begin{align}
\iota_{[a}\Omega^{(1)} \cdot  \frac{\d}{\d v}\big(\iota_{b]}\Omega^{(1)}\big)&=-16\sin^4\theta\,\, e^{2D}(P^{(1)}_{\bf 7})_{ab}{}^{cd}(\cald\omega+W\cald\beta-h_{\ul v})_{cd} \\
&+\sqrt{2}\sin^2\theta\, e^{2D+\phi}\Phi\cdot (\gamma_{ab} f_{\ul v}) \, , \cr
\iota_{[a}\Omega^{(2)} \cdot \frac{\d}{\d v}\big(\iota_{b]}\Omega^{(2)}\big)&=-16\sin^4\theta\,\, e^{2D}(P^{(2)}_{\bf 7})_{ab}{}^{cd}(\cald\omega+W\cald\beta+h_{\ul v})_{cd}\cr
&+\sqrt{2}\sin^2\theta\, e^{2D+\phi}\Phi\cdot (  f_{\ul v}\gamma_{ab}) \, . \nonumber
\end{align}

\section{Derivation of the general ansatz of Section \ref{genans}}
\label{app:genansatz}

In this appendix we show how the ansatz we use in Section \ref{genans}  to describe bound states of  D1-D5-P branes is the most general one compatible
with the BPS equations of \ref{summary}   under some hypothesis we will discuss  below.

\vspace{0.2cm}

We want to describe the back-reaction of bound states  of D1-D5-P in $\mathbb{R}^{1,1}\times Y \times T^4$.  The two assumptions we make are that the backgrounds are homogeneous and isotropic on $T^4$. Therefore all the fields in the ansatz  only depend
on the $(u,v,x^i)$ coordinates of  $\mathbb{R}^{1,1}\times Y$ and we  impose that $H$ has legs just along   $\mathbb{R}^{1,1}\times Y$ and 
that the RR-flux polyform $F$  splits as follows
\be\label{Ftotan}
F_{\rm tot}=F+ \hat{\rm vol}_{4}\wedge \tilde F\,,
\ee
where $F$ and $\tilde F$ have legs along $\mathbb{R}^{1,1}\times Y$ only.
The ten-dimensional self-duality condition reduces to the six-dimensional conditions
\be\label{6SD}
\tilde F=e^{4 \hat G} *_6 \lambda(F) \, ,
\ee
where $*_6$ uses the complete, warped, six-dimensional metric.  Moreover $F$ and $\tilde F$ must satisfy the following six-dimensional Bianchi identities/equations of
motion
\begin{subequations}\label{6BI0bis}
\begin{align}
\d_H F&=0 \, , \label{6BIbis}\\
\d_H\tilde F&=0 \, , \label{6BI2bis}
\end{align}
\end{subequations}
where  $\d_H\equiv \d - H$.  In this Appendix we will  adapt the flux decomposition \eqref{fluxdec} to the metric \eqref{metric244}. 
With obvious notation, we can write
\begin{subequations}
\begin{align}
\label{6dflux1}
& F=E^{\ul u}\wedge E^{\ul v}\wedge f_{\ul{uv}}+E^{\ul v}\wedge f_{\ul v}+E^{\ul u}\wedge f_{\ul u}+f  \, , \\
\label{6dflux2}
& \tilde F=E^{\ul u}\wedge E^{\ul v}\wedge \tilde f_{\ul{uv}}+E^{\ul v}\wedge\tilde f_{\ul v}+E^{\ul u}\wedge \tilde f_{\ul u}+\tilde f \, .
\end{align}
\end{subequations}
Notice that here  $f_{\ldots}$  denote the flux components along the four-dimensional space $Y$ and not, as in the rest of the paper, the eight-dimensional
components transverse to $(u,v)$. \\

The duality condition (\ref{6SD}) splits into the following set of duality conditions on $Y$:
\be
\label{dualityeqs}
\tilde f_{\ul{uv}}=e^{4 \hat G}\tilde *_4 \lambda(f) \,,\quad~ \tilde f_{\ul{v}}=e^{4 \hat G}\tilde *_4 \lambda(f_{\ul{v}} )\,,\quad~ \tilde f_{\ul{u}}=-e^{4 \hat G}\tilde *_4 \lambda(f_{\ul{u}} )\,, \quad~ \tilde f=e^{4 \hat G}\tilde *_4 \lambda(f_{\ul{uv}}) \,,
\ee
where $\tilde *_4$ uses the complete warped four-dimensional metric $e^{2G}\d s^2_4$. 

\vspace{0.2cm}

We will now  solve the BPS conditions summarised in Section \ref{summary}.

\subsection{Equation  \eqref{susysum1a}}

Let us start with \eqref{susysum1a}
\beq
\label{susysum1abis}
\d \chi =  \iota_K H \, .
\eeq
As already discussed in Section \eqref{projections},  the most general local solution is provided by\footnote{More precisely, the most general  local solution takes the form $H=\d B$, with $B$ $u$-independent and such that $\iota_KB=-\chi+\d\lambda$ where  $\lambda$ is a function. On the other hand, we can perform a gauge transformation $B\rightarrow B+\d (\lambda\d u)$ which allows to bring $\iota_K B$ into the form $\iota_KB=-\chi$.}
\be
\label{Bsol}
B= -  e^{2D}\cos2\theta\,(\d u+\omega) \wedge(\d v+\beta)+ b \wedge  (\d v+\beta) + \calb\,,
\ee
where $b=b_i\d x^i$ and $\calb =\frac12 \calb_{ij}\d x^i\wedge \d x^j$ have legs just along $Y$.
Then
\be\label{Hsol}
\begin{aligned}
H=&-(\d u+\omega)\wedge (\d y+\beta)\wedge\big[\cald(e^{2D}\cos2\theta)-e^{2D}\cos2\theta\dot\beta\big]\\
&+(\d v+\beta)\wedge\big[\dot\calb-e^{2D}\cos2\theta\,\cald\omega-(\cald b-\dot\beta\wedge b)\big]\\
&+(\d u+\omega)\wedge (e^{2D}\cos2\theta\,\cald\beta)+\cald\calb-\cald\beta\wedge b \, .
\end{aligned}
\ee

\subsection{Equation \eqref{susysum1b}} 

We then pass to \eqref{susysum1b}. The isotropy condition (\ref{Ftotan})  implies that the `non-isotropic' components of $\d_H(e^{-\phi}\Psi)$ should vanish. Explicitly, we have
\be
\d_H(e^{-\phi}\Psi)|_{\text{non-is.}}=-\d_H\Big[e^{2(D+G+\hat G)-\phi}\sin2\theta\, (\d v+\beta)\wedge \sum^3_{A=1} J_A\wedge \hat J_A\Big] \, .
\ee  
We can use the gauge freedom (\ref{warpgauge}) to fix
\be\label{warpfix}
e^{-2 G}=e^{2(D+ \hat G)-\phi}\sin2\theta \, .
\ee
With this choice, by using (\ref{Hsol}) we have
\be
\begin{aligned}
\d_H(e^{-\phi}\Psi)|_{\text{non-is.}}=
&\sum^3_{A=1}\hat J_A \wedge \Big\{(\d v+\beta)\wedge(\cald J_A-\dot\beta\wedge J_A)-\cald\beta\wedge J_A\\
&+(\d u+\omega)\wedge(\d v+\beta)\wedge \big[e^{2D}\cos2\theta\,\cald\beta\wedge J_A\big]\Big\} \, . 
\end{aligned}\ee
Hence, $\d_H(e^{-\phi}\Psi)|_{\text{non-is.}}=0$  gives the  conditions
\begin{subequations}\label{iscond}
\begin{align}
\cald J_a-\dot\beta\wedge J_A&=0\,,\\
\cald\beta\wedge J_A&=0\quad~~~~~~\Leftrightarrow\quad~~~~*_4 \cald\beta=\cald\beta \, , \label{betasdapp}
\end{align}
\end{subequations}
where now $*_4$ denote the four-dimensional Hodge star with respect to the metric without warp-factor. 
The first condition tells us that the non-trivial $v$-dependence of the background constitutes a potential obstruction to the integrability of the almost hyperk\"ahler structure on $Y$.

\vspace{0.2cm}

Once we have imposed (\ref{warpfix}) and (\ref{iscond}), the remaining equations contained in (\ref{susysum1b}) split as follows
\begin{subequations}
\begin{align}
\d_H(e^{-\phi} \varphi )&=\iota_KF+\chi\wedge F \, , \label{splitF1}\\
\d_H(e^{4\hat G-\phi}\varphi)&=\iota_K\tilde F+\chi\wedge \tilde F \, , \label{splitF2}
\end{align}
\end{subequations}
where $\varphi$ is  the part of the  $\Psi$  with all legs in  $\mathbb{R}^{1,1}\times Y$:
\be
\varphi=e^{2D}\sin2\theta(\d v+\beta)\wedge (1+e^{4 G} {\rm vol}_4) \, .
\ee

Using (\ref{susysum1abis}) and $\mathcal{L}_k B=0$,  \eqref{splitF1} can be rewritten as 
\be
\d(e^{-\phi}e^B\varphi)=\iota_K(e^BF)
\ee
and can therefore be solved in the same way as \eqref{susysum1abis}. Locally we can write $F$ as 
\be
F=\d_H C=e^{-B}\d(e^BC) \, , 
\ee
where  $C=C_0+C_2+C_4$ is $u$-idenpendent.  $C$ must satisfy $\iota_K(e^BC)=-e^{-\phi}e^B\varphi$ and then 
\be
\iota_K C=-e^{-\phi}\varphi-\chi\wedge C \, , 
\ee
which means that we can set
\be
\begin{aligned}\label{RRpotsol1}
C&=-(\d u+\omega)\wedge (e^{-\phi}\varphi+\chi\wedge \calc)+c \wedge (\d v+\beta)+\calc\\
&=-e^{2D}(\d u+\omega)\wedge (\d v+\beta)\wedge [e^{-\phi}\sin2\theta+\cos2\theta\, \calc]+ c \wedge (\d v+\beta)+\calc\,,
\end{aligned}
\ee
where $c\equiv c_1+c_3$ and $\calc\equiv \calc_0+ \calc_2+ \calc_4$ are polyforms with legs along $Y$ and in the second step we have omitted the piece of $\varphi$ containing ${\rm vol}_4$  since it is irrelevant. 
(\ref{RRpotsol1}) provides a parametrization of the most general local solution of (\ref{splitF1}). 

The most general solution of (\ref{splitF2})  can be obtained in an almost identical way. Setting locally $\tilde F=\d_H\tilde C$,  
with $\tilde C= \tilde C_0+\tilde C_2+\tilde C_4$ gives the general local solution
\be
\begin{aligned}\label{RRpotsol2}
\tilde C&=-(\d u+\omega)\wedge (e^{4\hat G-\phi}\varphi+\chi\wedge \tilde \calc)+ \tilde c \wedge (\d v+\beta)+\tilde \calc\\
&=-e^{2D}(\d u+\omega)\wedge (\d v+\beta)\wedge \big(e^{4\hat G-\phi}\sin2\theta+\cos2\theta\, \tilde\calc\big) + \tilde c \wedge (\d v+\beta) +\tilde \calc\,,
\end{aligned}
\ee
where again $\tilde c\equiv\tilde c_1+\tilde c_3$ and $\tilde \calc\equiv\tilde \calc_0+\tilde \calc_2+ \tilde \calc_4$  have legs along $Y$,   and,   in the last line,
 we have omitted an irrelevant six-form.

\vspace{0.2cm}
Now that we found  two independent general solutions for (\ref{6BIbis}) and (\ref{6BI2bis}),  we have to compute $F=\d_HC$ and $\tilde F=\d_H \tilde C$ and impose the self-duality condition (\ref{6SD}).  In the notation of \eqref{6dflux1} and \eqref{6dflux2}, we have for $F$
\be
\label{f6d}
\begin{aligned}
f_{\ul{uv}}=&-e^{-2D}\cald(e^{2D-\phi}\sin2\theta)-\cos2\theta(\cald \calc-\cald\beta\wedge c) \\
&+e^{-\phi}\sin2\theta\,\dot\beta+(\cald \calb -\cald\beta\wedge b)\wedge (e^{-\phi}\sin2\theta+\cos2\theta\,\calc ) \, , \\
f_{\ul v}=&\,e^{-2D}\big[\dot \calc -\dot{\calb  } \wedge\calc-e^{2D-\phi}\sin2\theta\,(\cald\omega+W\cald\beta)+(\cald c-\dot\beta\wedge c)\\
&-(\cald b-\dot\beta\wedge b)\wedge \calc -(\cald \calb -\cald\beta\wedge b)\wedge c\big] \, , \\
f_{\ul u}=&e^{2D-\phi}\sin2\theta\,\cald\beta \, , \\
f=&\,(\cald \calc -\cald\beta\wedge c)-(\cald \calb -\cald\beta\wedge b)\wedge \calc \, , 
\end{aligned}
\ee
and for $\tilde F$ 
\be
\label{tildef6d}
\begin{aligned}
\tilde f_{\ul{uv}}=&-e^{-2D}\cald(e^{2D+4\hat G-\phi}\sin2\theta)-\cos2\theta(\cald\tilde \calc -\cald\beta\wedge \tilde c)\\
&+e^{4\hat G-\phi}\sin2\theta\,\dot\beta+(\cald \calb -\cald\beta\wedge b)\wedge (e^{4\hat G-\phi}\sin2\theta+\cos2\theta\,\tilde \calc) \, , \\
\tilde f_{\ul v}=&\,e^{-2D}\big[\dot{\tilde \calc }+\dot{\calb}  \wedge\tilde\calc-e^{2D+4\hat G-\phi}\sin2\theta\,(\cald\omega+W\cald\beta)+(\cald\tilde c-\dot\beta\wedge \tilde c)\\
&-(\cald b-\dot\beta\wedge b)\wedge\tilde \calc -(\cald\calb  -\cald\beta\wedge b)\wedge\tilde c\big] \, , \\
\tilde f_{\ul u}=&e^{2D+4G-\phi}\sin2\theta\,\cald\beta \, , \\
\tilde f=&\,(\cald\tilde \calc -\cald\beta\wedge \tilde c)-(\cald \calb -\cald\beta\wedge b)\wedge\tilde \calc \, .
\end{aligned}
\ee

\subsection{Equation \eqref{susysum2a} and  \eqref{susysum2b} }

Let us start from (\ref{susysum2a}). Using the duality conditions \eqref{dualityeqs} and the expressions for the fluxes \eqref{f6d}, \eqref{tildef6d},  
this becomes 
\be\label{susysum2abis}
\cald (2 \hat G-\phi)+\frac{1}{2}\,e^\phi\,\frac{\cos2\theta}{\sin2\theta}\, (\cald\calc_0 + e^{-4 \hat G} \cald \tilde \calc_0) =0\, .
\ee
To arrive at this expression we have rewritten the right-hand side  of (\ref{susysum2a}) as
\be
\begin{aligned}
\frac1{\sqrt{2}\,\sin^2(2\theta)}e^\phi\, (f\cdot\iota_a\Phi)\,\d x^a &= \frac{e^\phi}{2\,\sin2\theta}(\tilde*_4 f_3 + e^{-4 \hat G}\tilde*_4 \tilde f_3) \\
&= - \frac{e^\phi}{2\,\sin2\theta}(e^{-4 \hat G}\tilde f_{\ul{uv}1} + f_{\ul{uv}1}) \,.
\end{aligned}
\ee

Let us now look at \eqref{susysum2b}.  The r.h.s. is
\be
\frac1{\sqrt{2}}e^\phi\, \iota_a f\cdot\Phi = \frac{\sin2\theta}{2}\,e^\phi\,(f_1 + e^{-4 \hat G}\,\tilde f_1)\,,
\ee
and hence, substituting the expressions for $f$ and $\tilde f$, that equation reduces to
\be\label{susysum2bbis}
\cald \cos2\theta = \frac{1}{2} \sin2\theta\,e^\phi\,(\cald \calc_0 + e^{-4 \hat G} \cald \tilde \calc_0)\,.
\ee

Combining (\ref{susysum2abis}) and (\ref{susysum2bbis}) one obtains
\be
\cald (2 \hat G - \phi - \log \sin2\theta)=0\,,
\ee
which implies
\be
2 \hat G - \phi - \log \sin2\theta=0\, , 
\ee
up to an irrelevant constant which can be reabsorbed by a redefinition of the dilaton. The relation above and the gauge choice (\ref{warpfix}) leave only two independent quantities out of the three warp factors  $D$, $G$, $\hat G$ and the dilaton $\phi$.  In the following, we choose to keep as independent quantities $G$ and $\hat G$, so that 
\be
e^\phi = \frac{e^{2 \hat G}}{\sin2\theta}\,,\quad e^D=\frac{e^{- G}}{\sin2\theta}\,.
\ee

The remaining two scalars, $C_0$ and $\tilde C_0$, are not completely fixed at this point but are related by the relation (\ref{susysum2bbis}), which can be rewritten as
\be\label{eqscalars1}
\cald \calc_0+ e^{-4 \hat G} \cald \tilde \calc_0= 2 e^{-2 \hat G}\,\cald \cos2\theta\,.
\ee

\subsection{Self-duality of the RR field strengths}

Let us now examine the duality relations (\ref{dualityeqs}).  First, notice that by using (\ref{betasdapp}) the third condition of (\ref{dualityeqs}) is automatically satisfied. The other 
conditions
\be
\tilde f_{\ul{uv}}=e^{4\hat G}\,\tilde*_4  \lambda(f )  \quad\,, \quad  \tilde f=e^{4\hat G}\, \tilde*_4  \lambda(f_{\ul{uv}})  \, , 
\ee
once expanded in forms of fixed degree, can be written as
\begin{subequations}
\begin{align}
\label{sd1}
\cald \calc_2 - \cald\beta\wedge c_1 =& e^{2(G - \hat G)}\,*_4[ (e^{-2 \hat G}\cos2\theta - \calc_0) \cald \tilde\calc_ 0\cr
&  +(e^{2 \hat G}\cos2\theta\,\calc_0+\sin^22\theta)(\dot{\beta}-2\cald(\hat G- G))] \,, \\
\label{sd2}
 \cald \tilde\calc_2 - \cald\beta\wedge \tilde c_1 =& e^{2(G + \hat G)}\,*_4[ (e^{2 \hat G}\cos2\theta - \tilde\calc_0)\cald \calc_0\cr
& +(e^{-2 \hat G}\cos2\theta\,\tilde\calc_0+\sin^22\theta)(\dot{\beta}+2\cald(\hat G+  G))] \,, \\
 \label{sd4}
 \cald \calb  - \cald\beta\wedge b =&*_4[ e^{2(G -\hat  G)}\cald  \tilde C_0 + e^{2  G} \cos2\theta (\dot{\beta}-2\cald (\hat G- G))]\,,  \\
 \label{sd3}
 \cald \calc_0 - e^{-4 \hat G}\cald \tilde\calc_0  =& -4e^{-2 \hat G}\cos2\theta\,\cald \hat G\, .
\end{align}
\end{subequations}
The last equation, \eqref{sd3},  provides  another  relation on the scalars in the ansatz, which,  combined  with (\ref{eqscalars1}) gives
\be
\cald (\calc_0 - e^{-2\hat G}\cos2\theta)=0\,,\quad \cald (\tilde\calc_0 - e^{2\hat G}\cos2\theta)=0
\ee 
and hence, up to a constant,
\be
\label{Cscalars}
\calc_0= e^{-2\hat G}\cos2\theta\,,\quad \tilde\calc_0= e^{2\hat G}\cos2\theta\,.
\ee
These relations can be used to simplify the previous ones
\be\label{eqsZs}
\begin{aligned}
& \cald \calc_2 - \cald\beta\wedge c_1= *_4 [e^{2(G -\hat G)}\dot{\beta}+\cald e^{2(G -\hat G)}] \, , \\
& \cald \tilde\calc_2 - \cald\beta\wedge \tilde c_1=*_4 [e^{2(G +\hat G)}\dot{\beta}+\cald e^{2(G +\hat G)}] \, , \\
& \cald \calb - \cald\beta\wedge b= *_4 [e^{2G}\cos2\theta\,\dot{\beta}+\cald (e^{2G} \cos2\theta)] \, .
\end{aligned}
\ee

\vspace{0.2cm} 
Let us now consider the last non-trivial duality constraint  in (\ref{dualityeqs})
\be
 \tilde f_{\ul{v}}=e^{4G} \tilde *_4 \lambda(f_{\ul{v}})  \, . 
 \ee
To this extent it is convenient to introduce some more notation,  the three two-forms
\be
\begin{aligned}
& \Theta= \dot{\calc}_2 + \cald c_1-\dot{\beta}\wedge c_1 \,, \\ 
& \tilde\Theta=\dot{\tilde\calc}_2 + \cald\tilde c_1-\dot{\beta}\wedge \tilde c_1\,  , \\
& \Theta_b = \dot{\calb}+\cald b - \dot{\beta}\wedge b\, .
\end{aligned}
\ee
Then the self-duality constraints read
\be
\begin{aligned}
 \frac{\d}{\d v} (e^{2\hat G}\cos2\theta)=&e^{4(\hat G-G)}*_4 [\dot{\calc}_4-\Theta_b\wedge \calc_2+(\cald c_3 - \dot{\beta}\wedge c_3)\\
 & \quad - (\cald \calb - \cald\beta\wedge b)\wedge c_1]\,, \\
 \frac{\d}{\d v} (e^{-2\hat G}\cos2\theta)=&e^{-4(\hat G+G)}*_4 [\dot{\tilde\calc}_4-\Theta_b\wedge \tilde\calc_2+(\cald \tilde c_3 - \dot{\beta}\wedge \tilde c_3)\\
 & \quad - (\cald \calb - \cald\beta\wedge b)\wedge \tilde c_1]\,, 
\end{aligned}
\ee
 and
 \be
\label{eqomega}
 \cald \omega + *_4 \cald\omega + 2 W \,\cald\beta =e^{2(G-\hat G)}\tilde\Theta+ e^{2(G+\hat G)}*_4 \Theta -e^{2 G}\cos2\theta\, (\Theta_b+*_4 \Theta_b)\,.
\ee

\subsection{Equation \eqref{susysum2c}}

Using the expression for the fluxes \eqref{f6d} and the self-duality conditions, the supersymmetry constraint  \eqref{susysum2c} becomes
\be
\begin{aligned}
\label{susysum2cbis}
\frac{\d}{\d v}(\cos 2\theta)  =&   \frac{\sqrt{2}}4e^{2D+\phi} f_{\ul v}\cdot\Phi   \\
=& \frac{1}{2}\sin2\theta\,e^{\phi}(\dot{\calc}_0+  e^{-4\hat G}\dot{\tilde\calc}_0 ) \, ,
\end{aligned}
\ee
which is identically satisfied by the relations \eqref{Cscalars}.

\subsection{Equations \eqref{susysum2d} and  \eqref{susysum2e} } 

The last equations to analyse are  \eqref{susysum2d} and  \eqref{susysum2e}. 
As shown in Section \ref{projections},  the requirement of having supersymmetric D1-D5-P systems forces the two eight-dimensional spinors $\eta_1$ and $\eta_2$
to be proportional
\be
\eta_1=2^{1/4}\, \sin\theta\, \eta\qquad\,,\qquad\eta_2=2^{1/4}\, \cos\theta\, \eta \, .
\ee

Thus the two Spin(7) structures characterizing the most general background with null $K$ collapse to a single one
\be
\Omega^{(1)}=\sqrt{2}\sin^2\theta\,\Omega \qquad \,,\qquad \Omega^{(2)}=\sqrt{2}\cos^2\theta\,\Omega
\ee
associated with the four-form \eqref{Omegared}
\be
\Omega =  e^{4 G}{\rm vol}_4+e^{4 G} \hat{{\rm vol}}_{4}-e^{2G+2 \hat G}\sum^3_{A=1}J_A\wedge \hat J_A \, ,
\ee
where $J_A$ and $\hat J_A$ define  almost hyperK\"ahler structures on $Y$ and $T^4$ respectively (see \eqref{hyperkahlerstr}). 

This brings some simplifications to  \eqref{susysum2d} and  \eqref{susysum2e}.  In particular, their sum and difference give
\begin{subequations}
\begin{align}
\label{susysum2debis}
& 16\sqrt{2}(P_{\bf 7})_{ab}{}^{cd}(h_{\ul v})_{cd} = -e^{\phi}\Phi\cdot (\frac{1}{\sin^2\theta}\,\gamma_{ab}f_{\ul v}-\frac{1}{\cos^2\theta}\,\ \, f_{\ul v}\gamma_{ab}) \, , \\
& \iota_{[a}\Omega\,\cdot \,\frac{\d}{\d v}\big(\iota_{b]}\Omega\big)+8\, e^{2D}(P_{\bf 7})_{ab}{}^{cd}(\cald\omega+W\cald\beta)_{cd} 
= \frac{e^{2D+\phi}}{2\sqrt{2}}\Phi\cdot \Big(\frac{1}{\sin^2\theta}\,\gamma_{ab}f_{\ul v}+\frac{1}{\cos^2\theta}\,f_{\ul v}\gamma_{ab}\Big) \, . \nonumber
\end{align}
\end{subequations}

 \bigskip
 
 In order to simplify further the equations above, we need some extra work. Let us first introduce the   two-form
\be
\psi\equiv \sum_A\psi_A J_A=\frac{1}{4}\epsilon^{ABC}(J_A\cdot\dot J_B)J_C\equiv \frac{1}{8}\epsilon^{ABC}(J_A)^{ij}(\dot J_B)_{ij}J_C \,,
\ee
which is anti-self-dual on $Y$.
Notice that,  for any self-dual   and  anti-self-dual 2-forms   $\rho_{\rm sd}$ and $\rho_{\rm asd}$  we have
\be
(J_A)_{[i}{}^{ k}(\rho_{\rm sd})_{j]k}\equiv 0 \qquad\,, \qquad \rho_{\rm asd}\equiv \frac12\sum_A(J_A\cdot \rho_{\rm asd})J_A\,.
\ee
These identities can be used to check that  $\psi$ can be rewritten as 
\be
\psi = - \frac{1}{4} \sum_A(J_A)_{i}{}^k  (\dot J_A)_{jk}\d x^i\wedge \d x^j \, . 
\ee
Then,  it is tedius but straightforward to prove that 
\begin{subequations}
\label{uglyid}
\begin{align}
\iota_{[m}\Omega\cdot \frac{\d}{\d v}\iota_{n]}\Omega&=-4e^{2\hat G}\sum_A\psi_A\, (\hat J_A)_{mn}   \,,\\
\iota_{[i}\Omega\cdot \frac{\d}{\d v}\iota_{j]}\Omega&=-4e^{2G}\psi_{ij} \,,   \\
(P_{\bf 7})_{ab}{}^{ij}(\cald\omega+W\cald\beta)_{ij}&=\frac14e^{2\hat G-2 G}\sum_A(J_A\cdot\cald\omega)(\hat J_A)_{ab} \,, \\
(P_{\bf 7})_{ij}{}^{hk}(\cald\omega+W\cald\beta)_{hk}&=\frac14[(1-*_4)\cald\omega]_{ij} \,,  \\
(P_{\bf 7})_{ab}{}^{ij}(h_{\ul{v}})_{ij}&=-\frac14e^{2G-2\tilde G}\sum_A(J_A\cdot h_{\ul{v}})(\hat J_A)_{ab} \,, \\
(P_{\bf 7})_{ij}{}^{hk}(h_{\ul{v}})_{hk}&=-\frac14[(1-*_4)h_{\ul{v}}]_{ij} \,,  \\
\Phi\cdot(\gamma_{ab}f^{\rm tot}_{\ul v})&=-\sqrt{2}\,\sin2\theta\, e^{2\hat G-2 G}\sum_A(J_A\cdot f_{\ul{v}2})(\hat J_A)_{ab}\,, \\
\Phi\cdot(f^{\rm tot}_{\ul v}\gamma_{ab})&=-\sqrt{2}\,\sin2\theta\, e^{2\hat G-2 G}\sum_A(J_A\cdot f_{\ul{v}2})(\hat J_A)_{ab} \,, \\
\Phi\cdot(\gamma_{ij}f^{\rm tot}_{\ul v})&=-\sqrt{2}\,\sin2\theta\, [(1-*_4)f_{{\ul v}2}]_{ij} \,, \\
\Phi\cdot(f^{\rm tot}_{\ul v}\gamma_{ij})&=-\sqrt{2}\,\sin2\theta\, [(1-*_4)f_{{\ul v}2}]_{ij}  \,,
\end{align}
\end{subequations}
where, on the left-hand side, we introduced the superscript ${}^{\rm tot}$  to distinguish  the  eight-dimensional RR fluxes  defined in \eqref{fluxdec} from the
four-dimensional ones in  (\ref{6dflux1}).

 \vspace{0.2cm}
 
Using the identities \eqref{uglyid} and the expression for the fluxes derived in the previous sections,
 it is possible to reduce the  constraint \eqref{susysum2debis} to a more readable form
\begin{subequations}
\begin{align}
\label{susysum2debis1}
(1-*_4) \Theta  &=2\, e^{2G-2\hat G}\psi \,,\\ 
 (1-*_4) \Theta_b &=2\cos2\theta\, e^{2G}\,\psi  \, .
\end{align}
\end{subequations}
 Finally, combining the previous equations with self-duality condition \eqref{eqomega}  gives
\be
 (1-*_4) \tilde\Theta=2\, e^{2G+2\hat G}\,\psi \, .
\ee

\subsection{Summary and relation with the notation of previous works}
 \label{app:summary}
 
In this  section, we summarize  our ansatz and the complete set of supergravity equations one has to solve to find supergravity solutions describing D1-D5-P systems.
We change the notation to make it  compatible with previous works  \cite{Gutowski:2003rg,Cariglia:2004kk,Bena:2007kg,Bena:2011dd,Giusto:2012gt,Giusto:2012jx}.
We rename the various metric and gauge field coefficients as follows:
\be
\begin{aligned}
&Z\to Z_1\,,\quad \tilde Z\to Z_2\,,\quad Z_b\to Z_4\,,\quad W\to \frac{\mathcal{F}}{2}\,,\\
&c_1\to a_1\,,\quad \tilde c_1\to a_2\,,\quad b\to a_4\,,\\
&  \calc_2 \to  \gamma_2\,,\quad  \tilde \calc_2 \to \gamma_1\,,\quad \calb \to \delta_2\,,\quad c_3\rightarrow x_3\,.
\end{aligned}
\ee
Note that now the number subscripts do not denote anymore the form degree but rather refer to the 
particular multiplet the corresponding fields belong to. 

\vspace{0.2cm}

In the new notation the ansatz for the metric in string frame, the dilaton, the NSNS B-field and the RR gauge fields is
 \be\label{ansatzsummary}
 \begin{aligned}
\d s^2_{(10)} &= -\frac{2\alpha}{\sqrt{Z_1 Z_2}}\,(\d v+\beta)\,\Big[\d u+\omega + \frac{\mathcal{F}}{2}(\d v+\beta)\Big]+\sqrt{Z_1 Z_2}\,\d s^2_4+\sqrt{\frac{Z_1}{Z_2}}\,\d \hat{s}^2_{4}\, ,\\
e^{2\phi}&=\alpha\,\frac{Z_1}{Z_2}\, ,\\
B&= -\frac{\alpha\,Z_4}{Z_1 Z_2}\,(\d u+\omega) \wedge(\d v+\beta)+ a_4 \wedge  (\d v+\beta) + \delta_2\,,\\ 
 C_0&=\frac{Z_4}{Z_1}\, ,\\
C_2 &= -\frac{\alpha}{Z_1}\,(\d u+\omega) \wedge(\d v+\beta)+ a_1 \wedge  (\d v+\beta) + \gamma_2\,,\\ 
C_4 &= \frac{Z_4}{Z_2}\, \hat{\mathrm{vol}}_{4} - \frac{\alpha\,Z_4}{Z_1 Z_2}\,\gamma_2\wedge (\d u+\omega) \wedge(\d v+\beta)+x_3\wedge(\d v + \beta)\,,
\end{aligned}
\ee
where
\be
\alpha = \frac{Z_1 Z_2}{Z_1 Z_2 - Z_4^2}\,.
\ee

\vspace{0.2cm}

As already mentioned at the end of Section~\ref{genans}, the search of a solution can be systematized 
by solving the equations in the following order

 \begin{itemize}
\item Equations for $\d s^2_4, \beta$:
  \begin{subequations}\label{eqJbeta}
 \begin{align}\label{eqJ}
& \d J_A= \frac{\d}{\d v}(\beta\wedge J_A)\,,\quad *_4 J_A = - J_A\,,\quad J_A\wedge J_B = -2 \,\delta_{AB}\,\mathrm{vol}_4\,, \\
&\label{eqbeta} *_4\cald\beta=\cald\beta\,;
\end{align}
\end{subequations}

 \item Equations for $ Z_1, a_2, \gamma_1$:
 \begin{subequations}\label{eqZ1a2}
 \begin{align}\label{eqZ1}
& *_4(\cald Z_1 + \dot{\beta}\,Z_1)= \cald \gamma_1 - a_2\wedge \cald\beta\,,\\
&\label{eqTh2}
\Theta_2 - Z_1 \,\psi = *_4 (\Theta_2 - Z_1 \,\psi) \quad~~~~\mathrm{with}\quad \Theta_2 = \cald a_2 -\dot\beta\wedge a_2+ \dot \gamma_1\,;
\end{align}
\end{subequations}
with 
\be
\psi = \frac{1}{8}\,\epsilon^{ABC}(J_A)^{i j}  ( \dot{J}_B)_{ij}  J_C\, ;
\ee

\item Equations for $Z_2, a_1, \gamma_2$:
 \begin{subequations}\label{eqZ2a1}
 \begin{align}\label{eqZ2}
& *_4(\cald Z_2 + \dot{\beta}\,Z_2)= \cald \gamma_2 - a_1\wedge \cald\beta\,,\\
&\label{eqTh1}
\Theta_1 - Z_2 \,\psi = *_4 (\Theta_1 - Z_2 \,\psi) \quad~~~~\mathrm{with}\quad \Theta_1 = \cald a_1 -\dot\beta\wedge a_1+ \dot \gamma_2\,;
\end{align}
\end{subequations}

\item Equations for $Z_4, a_4, \delta_2$:
 \begin{subequations}\label{eqZ4a4}
 \begin{align}
& \label{eqZ4}
 *_4(\cald Z_4 + \dot{\beta}\,Z_4)= \cald \delta_2 - a_4\wedge \cald\beta\,,\\
&\label{eqTh4}
\Theta_4 - Z_4 \,\psi = *_4 (\Theta_4 - Z_4 \,\psi) \quad~~~~\mathrm{with}\quad \Theta_4 = \cald a_4 -\dot\beta\wedge a_4+ \dot\delta_2\,;
\end{align}
\end{subequations}

\item Equations for $\omega,\mathcal{F}$:
\begin{subequations}\label{eqcalFomega}
\begin{align}\label{eqomegas}
&\cald \omega + *_4  \cald\omega + \mathcal{F} \,\cald\beta = Z_1\, *_4 \Theta_1+ Z_2\,\Theta_2 -Z_4\, (\Theta_4 +*_4  \Theta_4)\,,\\
&\label{eqcalF} *_4\cald *_4 L+2\,\dot\beta_i\, L^i  +\frac14 \frac{Z_1 Z_2}{\alpha} \,\dot{g}^{ij} \dot{g}_{ij}-\frac12 \frac{\d}{\d v}\Bigl[ \frac{Z_1 Z_2}{\alpha}\,g^{ij} \dot{g}_{ij}\Bigr] \nonumber \\
&-\dot{Z_1} \dot{Z_2}-Z_1 \ddot{Z_2} - \ddot{Z_1} Z_2+(\dot{Z}_4)^2 + 2 \,Z_4 \ddot{Z}_4\nonumber \\
&+\frac{1}{2} *_4 \Big[(\Theta_1-Z_2\psi)\wedge (\Theta_2-Z_1\psi) - (\Theta_4-Z_4\psi)\wedge (\Theta_4-Z_4\psi)\nonumber \\
&\qquad + \frac{Z_1 Z_2}{\alpha}\, \psi\wedge \psi-2 \psi\wedge \cald\omega\Bigr]=0\,,
\end{align}
\end{subequations}
with
\be
L = \dot{\omega} + \frac{\mathcal{F}}{2}\,\dot{\beta}-\frac{1}{2}\,\cald \mathcal{F}\,;
\ee
\item Equation for $x_3$:
\be\label{eqx3}
\cald x_3 - \dot{\beta}\wedge x_3 - \Theta_4\wedge \gamma_2+ a_1 \wedge (\cald \delta_2- a_4\wedge \cald \beta) = Z_2^2\,*_4 \frac{\d}{\d v}\Bigl(\frac{Z_4}{Z_2}\Bigr)\,.
\ee
\end{itemize}
Only the first set of equations (\ref{eqJbeta}) is non-linear. The remaining conditions, if solved in the order presented above, represent a set of linear equations for their respective unknowns. 


\end{appendix}



\bibliographystyle{utphys}      
\providecommand{\href}[2]{#2}\begingroup\raggedright\endgroup


\end{document}